\documentclass[10pt,journal]{IEEEtran}
\usepackage{cite}
\usepackage{amsmath,amssymb,amsfonts}
\usepackage{algorithm}
\usepackage{algorithmic}
\usepackage{graphicx}
\usepackage{dcolumn}
\usepackage{bm}
\usepackage{textcomp}
\usepackage{xcolor}
\usepackage{qcircuit}
\usepackage{hyperref}
\usepackage[capitalise]{cleveref}
\usepackage{comment}
\usepackage{booktabs}
\usepackage{physics}
\usepackage{float}
\usepackage{amsthm}
\usepackage{subfigure}
\usepackage[english]{babel}
\usepackage[utf8]{inputenc}
\usepackage[T5,T1]{fontenc}
\usepackage{enumerate}
\usepackage{soul}
\usepackage{indentfirst}
\usepackage{setspace}
\usepackage{units}

\makeatletter
\def\fnum@figure{Fig.~\thefigure}
\makeatother

\crefname{equation}{}{}
\crefname{figure}{Fig.}{Figs.}
\crefname{table}{Table}{Tables}

\creflabelformat{equation}{(#2#1#3)}
\creflabelformat{figure}{#2#1#3}
\creflabelformat{table}{#2#1#3}

\crefname{algocf}{algorithm}{algorithms}
\Crefname{algocf}{Algorithm}{Algorithms}
\crefname{assumption}{assumption}{assumptions}

\newcommand{\N}{{N_{\rm CNOT}}}
\newcommand{\tN}{{\tilde N_{\rm CNOT}}}
\newcommand{\p}{p}
\newcommand{\pth}{{p_{\rm threshold}}}
\newcommand{\pnum}{{2\times 10^{-5}}}

\newcommand{\hs}{{\hat{\vb s}}}
\newcommand{\ts}{{\tilde{\vb s}}}
\newcommand{\ms}{{{\vb s}}}
\newcommand{\Arre}{{\vb A_{\rm rre}}}

\begin{document}

\title{Fault-tolerant noise guessing decoding of quantum random codes}

\author{
    Diogo Cruz, Francisco A. Monteiro,~\IEEEmembership{Member,~IEEE,} André Roque, Bruno C. Coutinho
    \thanks{This work was supported in part by the European Union’s Horizon 2020 Research and Innovation Program through the Project Quantum Internet Alliance (QIA) under Grant 820445; and in part by Fundação para a Ciência e Tecnologia / Ministério da Ciência, Tecnologia e Ensino Superior (FCT/MCTES), Portugal, through national funds and when applicable co-funded EU funds under projects UIDB/50008/2020 and 2022.05558.PTDC. Diogo Cruz was supported by FCT scholarship UI/BD/152301/2021.}
    \thanks{Diogo Cruz is with Instituto de Telecomunicações, and IST, Universidade de Lisboa, Lisbon, Portugal, e-mail: diogo.cruz@lx.it.pt.}
    \thanks{Francisco A. Monteiro is with Instituto de Telecomunicações, and ISCTE - Instituto Universitário de Lisboa, Lisbon, Portugal, e-mail: francisco.monteiro@lx.it.pt.}
    \thanks{André Roque is with Instituto de Telecomunicações, Lisbon, Portugal, e-mail: andrejvroque@tecnico.ulisboa.pt.}
    \thanks{Bruno Coutinho is with Instituto de Telecomunicações, Lisbon, Portugal, e-mail: bruno.coutinho@lx.it.pt.}
}

\maketitle

\begin{abstract}
This work addresses the open question of implementing fault-tolerant QRLCs with feasible computational overhead. We present a new decoder for quantum random linear codes (QRLCs) capable of dealing with imperfect decoding operations. A first approach, introduced by Cruz et al., only considered channel errors, and perfect gates at the decoder. Here, we analyze the fault-tolerant characteristics of QRLCs with a new noise-guessing decoding technique, when considering preparation, measurement, and gate errors in the syndrome extraction procedure, while also accounting for error degeneracy. Our findings indicate a threshold error rate ($\pth$) of approximately $\pnum$ in the asymptotic limit, while considering realistic noise levels in the mentioned physical procedures.
\end{abstract}


\section{Introduction}

It is known that classical random linear codes (RLCs) are capacity-achieving \cite{Duffy_Medard_TIT_2019}, however, until the advent of guessing random additive noise decoding (GRAND) their decoding was not practical, apart some decoders based on trellises (as pointed out in \cite{Cruz_Access_2023}). GRAND has been proposed with the aim of reducing end-to-end latency in coded wireless systems, which has been a drawback for a long time. The rationale in the original proposal of GRAND was that by using short codewords, the so-called interleavers, used to make the errors independent and identically distributed (iid), would no longer be required \cite{An_Medard_Duffy_2020}. Using short blocks in wireless systems also helps to better adapt to the channel variations when applying precoding techniques \cite{vucetic19, Alberto_CSNDSP_2020, Monteiro_2010}. In the quantum realm, due to technical limitations in manipulating qubits, short block codes appear as natural candidates for quantum error correction codes (QECCs) \cite{prevalent_error_types_JSAIT_2020,Shor1995,Steane1996,fowler_surface_2012,calderbankGoodQuantumErrorcorrecting1996}. These limitations also necessitate the development of fault-tolerant techniques to handle noise and errors in quantum operations \cite{GottesmanPhD}.

Likewise classical RLCs, quantum random linear codes (QRLCs) attain the capacity of the quantum channels, but no practical decoder existed for them until the advent of quantum guessing random additive noise decoding (QGRAND), which allowed to numerically assess their performance for the first time \cite{Cruz_Access_2023}. A recent work also used a GRAND-like approach to decode several families of structured quantum codes which are based on stabilizer codes \cite{Chandra_Hanzo_Access_2023}. QGRAND has also been applied to the purification of quantum links, taking advantage of the connection between purification and error correction \cite{Roque2023}, which will have great implications on the way routing is implemented in quantum networks \cite{Santos2023, bugalho2023, coutinho_robustness_2022,angkun_2020}.

QRLCs are a much more flexible solution than other structured quantum codes for QECCs, with advantages in respect to the state-of-the-art solutions designed to detect and correct errors in quantum setups \cite{GoogleQ_Nature_2023, Autoencoders_quantum_2023}. In contrast to structured codes, which may only exist for a very limited number of code rates and codeword lengths \cite{roffe2019quantum, Babar_Hanzo_2019}, QRLCs can exist for a wide range of coding rates and codeword lengths that may better fit some particular applications. A method to generate QRLCs efficiently was proposed in 2013 \cite{brown_short_2013}, however, almost no practical method existed until recently to decode them until the proposal in \cite{Cruz_Access_2023}.
The channel model used in that work was a Shannon-like channel, where errors occur only in the channel and all the decoding process is perfect. However, in all current technologies implementing qubits, the errors that take place in the quantum gates of the decoding circuit cannot be ignored. Hence, a practical challenge remained after \cite{Cruz_Access_2023}: can a QRLC-QGRANDf system be made practical in the presence of the extra errors coming from the quantum gates, enabling fault-tolerant QECCs based on QRLCs?

This paper shows that, surprisingly, due to the particular way that the syndrome extraction takes place in codes based on stabilizers, some heavy reduction of the effects of those errors takes place, making the whole system viable. Building on previous work \cite{Cruz_Access_2023,Roque2023}, we present a comprehensive analysis of fault-tolerant QRLCs, incorporating the effects of preparation, measurement, and gate errors. Our results show that QRLCs, decoded with the proposed method, exhibit robust error correction capabilities with a threshold error rate $\pth$ of approximately $\pnum$. This advancement paves the way for practical implementations of QRLCs in quantum error correction, contributing to the development of scalable and resilient quantum systems.

Although our results suggest that QGRAND could in theory enable a fault-tolerant implementation of QRLCs, some challenges remain that limit its usefulness in that regime. QGRAND is most suitable for situations where the noise entropy is relatively low, in which case decoding becomes computationally efficient. However, in the fault-tolerant regime where $n$ may be considered to be considerably large or it is necessary to iteratively apply error correction to suppress errors, the noise entropy can be considerably high. In this regime, the optimal procedure described in this paper becomes infeasible, and suboptimal heuristics would have to be introduced. Nonetheless, this work paves the way for applications of QGRAND whenever the considered noise types all have low entropy, which encompasses setups with realistic noise conditions.

This paper is organized as follows. In \Cref{sec:setup} we introduce the setup considered in the analysis, and in particular its noise model. In \Cref{sec:notation}, we define some useful error notation terms and set the notation used throughout the paper. \Cref{sec:decoding} presents the decoding method, extended form \cite{Cruz_Access_2023} to account for degenerate errors. In \Cref{sec:asymptotic}, we present an analysis of the codes' performance for various qubit counts, and in \Cref{sec:conclusion} we present some final thoughts on our results.

\section{Setup and noise model}\label{sec:setup}

%\onecolumn
\begin{table*}[t]
\caption{Notation summary.}
\label{tab:notation_summary}
\begin{tabular}{|c|p{10cm}|l|}
\hline
\textbf{Variable} & \textbf{Description} & \textbf{Relationships} \\
\hline
$U$ & Unitary encoding circuit for a quantum error-correcting code & \\
\hline
$S_i$ & The $i^{\text{th}}$ minimal stabilizer of the code ($1 \leq i \leq n-k$ ) & $S_i = UZ_{i+k}U^\dagger$ \\
\hline
$\bar{X}_j$ & Logical $X$ operator on the $j^{\text{th}}$ encoded qubit ($1 \leq j \leq k$) & $\bar{X}_j = UX_jU^\dagger$ \\
\hline
$\bar{Z}_j$ & Logical $Z$ operator on the $j^{\text{th}}$ encoded qubit ($1 \leq j \leq k$) & $\bar{Z}_j = UZ_jU^\dagger$ \\
\hline
$\mathcal{N}$ & List of all possible errors $E_i$ in the noise model, with associated probabilities $p_i$ & \\
\hline
$E_i^B$ & A base error (error affecting a single qubit or gate) & \\
\hline
$\omega$ & The number of base errors that compose a compound error & $E_j = E_{i_1}^B \cdots E_{i_\omega}^B$ \\
\hline
$e_i^l$ & Local error pattern corresponding to a base error $E_i^B$ & \\
\hline
$C, B$ & Unitary components of a syndrome extraction circuit, applied before/after an error $E_i^B$ occurs & $V = CB$ \\
\hline
$V_E$ & Syndrome extraction circuit affected by error $E_i^B$ & $V_E = Ce^l_iB$ \\
\hline
$e_i$ & Propagated error pattern on the main qubits after $E_i^B$ and subsequent circuit operations & $e_i = (Ce^l_iC^\dagger)_m$ \\
\hline
$e_i^s$ & Propagated error pattern on the ancilla qubits after $E_i^B$ and subsequent circuit operations & $e_i^s = (Ce^l_iC^\dagger)_a$ \\
\hline
$\vb A$ & Quantum check matrix of the code (binary representation in $[X|Z]$ format) & \\
\hline
$\vb e_i$ & Binary representation of the main error pattern $e_i$ (in $[Z|X]$ format) & \\
\hline
$\mathcal{L}$ & Logical error group generated by $\bar{X}_j$ and $\bar{Z}_j$ & \\
\hline
$\mathtt{L}_i$ & One of the $2^{2k}$ logical error patterns in $\mathcal{L}$ & $e_i = \mathtt{E}_i\mathtt{S}_i\mathtt{L}_i$ \\
\hline
$\mathcal{S}$ & Stabilizer group generated by $S_j$ & \\
\hline
$\mathtt{S}_i$ & One of the $2^{n-k}$ stabilizer patterns in $\mathcal{S}$ & $e_i = \mathtt{E}_i\mathtt{S}_i\mathtt{L}_i$ \\
\hline
$\mathtt{E}_i$ & Error pattern with the same syndrome as $e_i$ & $e_i = \mathtt{E}_i\mathtt{S}_i\mathtt{L}_i$ \\
\hline
$\vb 0$ & Vector representing zero syndrome ($n-k$ zero bits) & \\
\hline
$\hat{\vb s}_i$ & Syndrome associated with a (propagated) error pattern $e_i$ & $\hat{\vb s}_i = \vb e_i \vb A^T$ \\
\hline
$\mathcal{D}$ & A degenerate set: A set of error patterns with the same syndrome that can be corrected similarly & \\
\hline
$e_i^d$ & Representative of the errors in a degenerate set $\mathcal{D}$ & $e^d_i = \mathtt{E}_i \mathtt{L}_i$ \\
\hline
$g$ & Index of the syndrome extraction where an error $E_i^B$ occurred & \\
\hline
$\ts_i$ & Syndrome acquired in the same extraction as error $E_i^B$ & $\ts_i = \text{comp}_X(\text{bin}(e_i^s))$\\
\hline
$\hs_i$ & Syndrome acquired in a subsequent extraction after $E_i^B$ occurred & $\hs_i = \text{comp}_X(\text{bin}((Ve_iV^\dagger)_a))$ \\
\hline
$\ms_i$ & Measured syndrome in a particular syndrome extraction & \\
\hline
$s$ & List of all acquired syndromes over multiple extractions: $\{\vb{s}^1,\ldots,\vb{s}^q\}$ & \\ 
\hline
$\mathcal{Q}(E_j)$ & Syndrome sequence expected for a compound error $E_j$ & $\mathcal{Q}(E_j) = s_{i_1} \oplus \cdots \oplus s_{i_\omega}$ \\
\hline
\end{tabular}
\end{table*}
%\twocolumn

We use the same setup as in \cite{Cruz_Access_2023}, but consider the fault-tolerant regime, where the constituent quantum gates in the circuit may be affected by error. We consider an initial $k$-qubit $\ket{\psi}$ quantum state, to be encoded into $n>k$ qubits. \cite{brown_short_2013} presents a method of generating a random qubit encoding, which we use in this work. One starts by randomly selecting Clifford unitaries from the $\mathcal C_2$ group (i.e., Clifford unitaries for 2 qubits). There are $|\mathcal C_2|=11\,520$ such unitaries, and all of them can be built by simple combinations of the Hadamard ($H$), phase ($\sqrt{Z}$), and CNOT gates, which have efficient physical implementations in virtually any quantum setting \cite{fedorovQuantumComputingQuantum2022}. In matrix form, these are defined as
\begin{equation}
    H = \frac{1}{\sqrt{2}}\mqty(1 & 1 \\ 1 & -1),\; \sqrt{Z} = \mqty(1 & 0 \\ 0 & i),\;
	\text{CNOT} = \mqty( I &  0 \\  0 &  X),
\end{equation}
with $X,Y,$ and $Z$ the Pauli matrices, and $I$ the $2\times 2$ identity matrix.

After selecting these random unitaries from $\mathcal C_2$, one successively applies each of them to a random pair of qubits, taken from the set of $n$ qubits.

\begin{figure}[t]
    \includegraphics[width=\linewidth]{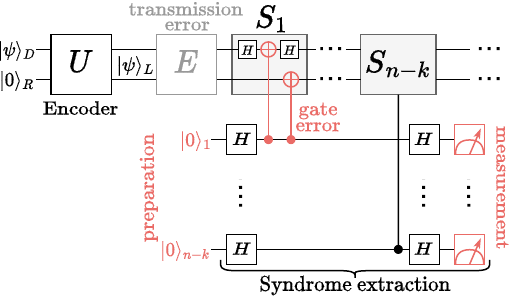}
    \caption{Noise model considered.}
    \label{fig:faulty_noise_model}
\end{figure}

This process leads to an encoding unitary for our stabilizer code which, when applied to the initial $k$ qubits and $(n-k)$ extra $\ket 0$ qubits added, returns a $n$-qubit encoded quantum state. As shown in \cite{brown_short_2013}, as long as $\order{n \log^2 n}$ gates are used, with a circuit depth of $\order{\log^3 n}$, the construction leads to a highly performant $(n,k)$ code, and from \cite{Gullans_Krastanov_Huse_Jiang_Flammia_2021} it is already known that these complexity orders can be further lowered.

We use these QRLCs to construct stabilizer codes. Compared to the approach in \cite{Cruz_Access_2023}, in this work, we consider a noise model that is more realistic by also including preparation, measurement, and gate errors. Given that, in practical applications, the error of 2-qubit entangling gates generally dominates over single-qubit gate errors \cite{fowler_surface_2012}, we focus on the former type of error. We assume that every gate in both the encoding and syndrome extraction steps is decomposed into the Clifford gates $\qty{\text{CNOT}, H, \sqrt{Z}}$.

For the noise statistics, we consider the model similar to the one in \cite{fowler_surface_2012}, but without single-qubit gate errors (see \cref{fig:faulty_noise_model}):
\begin{itemize}
    \item \textbf{CNOT gate errors:} After the ideal implementation of the CNOT$(a,b)$ gate, with qubit $a$ controlling $b$, it is assumed that one of the 15 errors of the form
    \begin{equation}
        O_aO_b \label{eq:error_set}
    \end{equation}
    with $O_a,O_b \in \qty{I, X, Y, Z}$ and excluding $O_aO_b \neq I_aI_b$, occurs with probability $p/15$. Here, $I$ is the identity gate, and $X,Y,Z$ are the Pauli matrices.
    \item \textbf{Preparation errors:} While setting the $(n-k)$ ancilla qubits (for each syndrome extraction) to $\ket 0$, each qubit has (independently) a probability $p$ of being prepared in the state $\ket 1$ instead.
    \item \textbf{Measurement errors:} While measuring each ancilla qubit to extract the syndrome, each measurement bit has a probability $p$ of being misread, so that a zero bit is read as a 1, and vice-versa. 
\end{itemize}
Unlike the model in \cite{Cruz_Access_2023}, to demonstrate the fault-tolerant properties of this model, we exclude a source of error between the encoding and syndrome extraction sections (i.e., the ``transmission error'' in \cref{fig:faulty_noise_model}), and instead focus on the case where the CNOT gate error stemming from the syndrome extraction dominates the noise statistics of the circuit. This simpler model facilitates the study of the QGRAND decoding approach in the fault-tolerant regime, which is the focus of this work. While possible (see \cref{sec:reducing_stabilizer_weight}), we make no further modifications to the circuit implementation.

\section{Error correction overview}\label{sec:notation}

In the fault-tolerant regime with a noisy gate model, degenerate errors play a significant role in the error correction capabilities of the code \cite{smith}. As a result, the approximation made in \cite{Cruz_Access_2023}, where codes were approximated to be non-degenerate, is no longer accurate, as it would significantly underestimate the code's capabilities.

Additionally, in the fault-tolerant regime, we must consider an iterated application of the syndrome extraction procedure, instead of a single application, in order to capably detect the errors being introduced by the syndrome extraction procedure itself. This is a common approach \cite{dennis_topological_2002,GoogleQ_Nature_2023} to quantum error correction when gate and measurement errors are non-negligible, and the decoding procedure has into account not just the syndrome from one extraction process, but the whole history of syndrome measurements.

\begin{figure}[t]
    \includegraphics[width=\linewidth]{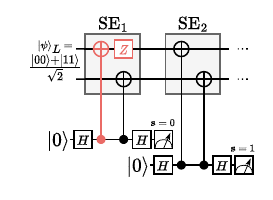}
    \caption{Simple example, with two syndrome extractions (SE). The error is not detected by the first syndrome extraction process, but it is detected by the second extraction.}
    \label{fig:model_relations}
\end{figure}

As a result of this added complexity, in this section we clarify the notation we use in this work. We use $\bar X_j, \bar Z_j$ to represent the logical operators corresponding to the unencoded operators $X_j, Z_j$, respectively. Given an encoding $U$ (see \cref{fig:faulty_noise_model}), the choice of minimal stabilizers $S_i$ and logical operators is not unique. Without loss of generality (W.l.o.g.), we consider the minimal stabilizer $S_i$ (for $1\leq i\leq n-k$) and logical operators $\bar X_j$, $\bar Z_j$ (for $1\leq j\leq k$) to be given by
\begin{align}
    S_i &= U Z_{i+k} U^\dagger\label{eq:default_stabilizers}\\
    \bar X_j &= U X_{j} U^\dagger\label{eq:default_logical_X}\\
    \bar Z_j &= U Z_{j} U^\dagger\label{eq:default_logical_Z}.
\end{align}

Following \cref{sec:setup}, the noise model enables us to create a list $\mathcal N = \qty{(p_0, E_0), (p_1, E_1), \ldots}$ of all the errors $E_i$ that the encoded quantum state may be subjected to, along with its respectively probability $p_i$ of occurring. 
An error $E_i$ refers to the qualitative process that occurred physically, such as ``a $X_2Z_3$ error occurred in the CNOT$(2,3)$ gate, and no other errors'', for example.

An error $E_i$ that corresponds to either only one wrongly prepared qubit, or one wrongly measured qubit, or one noisy CNOT gate, is called a \emph{base error}, and may be explicitly labeled as $E_i^B$. Every other error in the noise model of \cref{sec:setup} can be described as a combination of base errors. 

We consider the errors $E_i$ to be disjunctive, so only one error in $\mathcal N$ may occur, and their probability sums to 1. When using the base error notation $E_i^B$, we implicitly refer only to the specific base error that occurred, without making claims about the occurrence of other base errors. For example, using the noise model in \cref{sec:setup}, the base error $E^B=$``$X_2Z_3$ error in the CNOT$(2,3)$ gate'' would have a probability of occurring of $p/15$, while the corresponding error $E=$``$X_2Z_3$ error in the CNOT$(2,3)$ gate, and no other errors'' would have a probability of $(p/15)\times P(\text{no other base error occurs})$, which would possibly be much lower. We may use the shorthand notation $\hat E^B := E$ for errors where only one base error occurs.

Compound errors may be represented by their own symbol or as the product of errors that compose it. That is, for simplicity, given base errors $E_i^B$ and $E_j^B$, we also have the compound error notation 
\begin{equation}
    E_r = E_i^B \cap E_j^B \cap \qty(\bigcap_{m\neq i,j} \overline{E_m^B})=: E_i^BE_j^B
\end{equation}
An error $E_j$ is said to be of order $\omega$ if $\omega$ base errors suffice to describe it, so that $E_j = E_{i_1}^B \cdots E_{i_\omega}^B$. Note that the constituting base errors may stem from different syndrome extractions. 

Given a base error, let $e^l$ be the \emph{local error pattern} corresponding to the error that occurred locally. In the example above, we would have $e^l_i = X_2Z_3$. In general, for gate errors affecting only one CNOT gate, we would have an error pattern from \cref{eq:error_set}. For preparation and measurement errors, $e^l$ would be represented by $X$ operators in the appropriate ancilla qubits.

Unless the error $E_i^B$ occurs at the end of a syndrome extraction process, it will propagate through the rest of the quantum circuit, possibly impacting other qubits. Let $V$ be the unitary corresponding to one noiseless syndrome extraction process, minus the final measurement step of the ancilla qubits. We may partition $V$ into the unitaries $B$ and $C$, corresponding to the portion of the syndrome extraction circuit that occurs before and after the error $E_i^B$, respectively. If the syndrome extraction affected by $E_i^B$ is given by the unitary $V_E$, we have
\begin{align}
    V &= CB\\
    V_E &= Ce^l_iB = (e_i^s \otimes e_i) V.\label{eq:error_propagation}
\end{align}
The resulting (propagated) error pattern may affect non-trivially both the main $n$ qubits (main \emph{error pattern} $e_i$) and the $(n-k)$ ancilla qubits (\emph{ancilla error pattern} $e_i^s$). 

A Pauli string of $n$ qubits is an operator that is the product of the Pauli operators $X,Y$, and $Z$ on those qubits. It has the form
\begin{gather}
    e^{i\frac{\pi}{2}\phi}O_1O_2\cdots O_n,\\
    \text{ with }O_j \in \qty{X,Y,Z,I}
    \text{ and }\phi \in \qty{0,1,2,3}.\notag
\end{gather}
As $e^l_i$ is a Pauli string, and $C$ is a Clifford unitary, then both $e_i$ and $e_i^s$ are Pauli strings. Similarly, following \cref{eq:default_stabilizers,eq:default_logical_X,eq:default_logical_Z}, the minimal stabilizers and logical operators are also Pauli strings, since $U$ is a Clifford unitary as well. A Pauli string is said to have weight $t$ if it acts on $t$ qubits, that is, if its Pauli string contains $t$ Pauli operators (excluding the identity).

For the previous example with $E_i^B$, we would have $e^s_i \otimes e_i = C(X_2 Z_3)C^\dagger$. While the effect of $e^s_i$ is removed by the syndrome measurement, the same cannot be said of $e_i$. As $E_i^B$ propagates through the circuit, the pattern $e_i$ is picked up by subsequent syndrome extractions, and is ultimately the error pattern that our correction process needs to consider to undo the effect of $E_i^B$ on the main $n$ qubits. See \cref{fig:model_relations} for an example.

In \cref{fig:model_relations}, we showcase a simple example, with two syndrome extractions (SE), where there is only one minimal stabilizer $S_1 = X_1X_2$. An error occurs in the first CNOT gate, so $E=$ ``$Z_1 I_2$ error in first CNOT gate, and no other errors''. The error is not detected by the first syndrome extraction process, so $\ms=\ts=0$. By the end of the first syndrome extraction, the evolved uncorrected error is $e=Z_1I_2$. It is now detected by the second extraction, so $\ms=\hs=1$. If it is not corrected, subsequent extractions will behave similarly to the second one, returning the syndrome $1$.

Since $Y = iXZ$, any Pauli string of $n$ qubits can also be written in the form
\begin{equation}
    e^{i\frac{\pi}{2}\varphi}\qty(O_1^X\cdots O_n^X)\qty(O_1^Z\cdots O_n^Z),\label{eq:pauli_string}
\end{equation}
with $O_j^X \in \qty{X,I},\; O_j^Z \in \qty{Z,I}$ and $\varphi \in \qty{0,1,2,3}$.

The Pauli string may then be encoded as a binary row vector. In $[X|Z]$ format, it takes the form
\begin{gather}
    \mqty[b^X_1 & b^X_2 & \cdots & b^X_n & b^Z_1 & b^Z_2 & \cdots & b^Z_n],\label{eq:binary_repr}\\
    \text{with }b^P_j = \begin{cases}
        1, &\text{if } O_j = P\\
        0, &\text{if } O_j = I
    \end{cases}
    \text{ and }P\in\qty{X,Z}.
\end{gather}
In $[Z|X]$ format, the $b^X_j$ entries are swapped with $b^Z_j$. By default, binary vectors and matrices are represented in bold. We may use the functions
\begin{align}
    \text{bin}(e)&:=\vb e, \qquad \text{op}(\vb e):=e\label{eq:bin}\\
    \text{comp}_P(\vb e) &:= \mqty[b^P_1 & \cdots & b^P_c]\label{eq:comp},
\end{align}
with $P\in\qty{X,Z}$ and $c=|\vb e|/2$, to indicate the conversion to and from binary representation, and to refer to a particular component of $\vb e$, respectively. Calculations using binary are performed in $\mathbb F_2$, that is, using modular arithmetic mod 2. The functions bin and op stand for the transformations of Pauli operators from and to, respectively, binary arrays.

Let $\vb A$ be the quantum check matrix \cite{djordjevic_quantum_2021}, a $(n-k)\times 2n$ binary matrix (in $[X|Z]$ format) where each row $j$ encodes the minimal stabilizer $S_j$ of the code. This is a compact way of representing the encoding used. Let $\vb e_i$ be the binary representation of the error pattern $e_i$ as a $2n$-sized row vector, in $[Z|X]$ format. Any evolved error pattern $e_i$ can be written as
\begin{align}
    e_i = \mathtt{E}_i \mathtt{S}_i \mathtt{L}_i,\label{eq:error_decomposition}
\end{align}
where $\mathtt{L}_i$ is one of the $2^{2k}$ logical error patterns in the logical error group $\mathcal L$ generated by the logical operators $\bar X_j, \bar Z_j, 1\leq j\leq k$; $\mathtt{S}_i$ is one of the $2^{n-k}$ stabilizers in the stabilizer group $\mathcal S$ generated by the minimal stabilizers $S_j, 1\leq j \leq n-k$; and $\mathtt{E}_i$ is some error pattern with the same syndrome $\hat{\vb s}=\vb e_i \vb A^T$ as $e_i$ \cite{djordjevic_quantum_2021}. We use $\vb 0$ to represent the syndrome with all entries equal to zero. W.l.o.g., for the decomposition, we assume the phase factor $\varphi$ (see \cref{eq:pauli_string}) to be zero, since neglecting it adds at most a global phase to the encoded quantum state, which can be disregarded. As a result, we consider each error pattern to equal its inverse.

The decomposition in \cref{eq:error_decomposition} is not unique, and is dependant on the choice made for the particular logical operators, minimal stabilizers, and $\mathtt E_i$ patterns to use. For the sake of simplicity in the notation, in this work, it is assumed that such a decomposition is the unique one obtained deterministically by following the procedure described in \cref{sec:decoding}. Consequently, we assume that, associated with each error pattern $e_i$, there is a unique set of operators $\mathtt{E}_i, \mathtt{S}_i,$ and $\mathtt{L}_i$. In particular, for patterns with $\hs = \vb 0$, the operator $\mathtt E_i$ is the identity. Since all error patterns with the same syndrome will have the same error component $\mathtt E$, we use $\mathtt E_{\hs}$ to indicate the error component of the error patterns with syndrome $\hs$.

Compound error patterns, such as $e_r = e_ie_j$, may be easily encoded in binary form by using the modular sum (i.e. XOR), so that $\vb e_r = \vb e_i\oplus \vb e_j$ and $\hs_r = \hs_i\oplus \hs_j$. 
Given two error patterns $e_i, e_j$ with the same syndrome $\hat{\vb s}$, their product $e_r$ has syndrome $\hs_k = \hs\oplus \hs = \vb 0$, so $\mathtt E_r$ is the identity operator. Therefore, there is a unique $\mathtt{S}\in \mathcal S, \mathtt{L} \in \mathcal L$ such that $e_i = e_j \mathtt S \mathtt L$, with $\mathtt S = \mathtt S_i \mathtt S_j$ and $\mathtt L = \mathtt L_i \mathtt L_j$. 

A \emph{degenerate set} $\mathcal D$ is a set of evolved error patterns that can be treated similarly, for correction purposes. This set depends on $\hat{\vb s}$ and $\mathtt L$. Although all error patterns with the same syndrome $\hat{\vb s}$ have the same representative error pattern $\mathtt E_{\hat{\vb s}}$, not all can be corrected similarly. For that to be the case, their logical error component $\mathtt L$ must be the same. Since we know that two error patterns are degenerate if $e_ie_j \in \mathcal S$, we may verify this by computing $e_r = e_ie_j = (\mathtt E_i \mathtt S_i \mathtt L_i)(\mathtt E_j \mathtt S_j \mathtt L_j) = (\mathtt E_i \mathtt E_j) (\mathtt S_i \mathtt S_j) (\mathtt L_i \mathtt L_j) = \mathtt E_r \mathtt S_r \mathtt L_r$. For $e_r$ to be in $\mathcal S$, we must have $\mathtt E_r=\mathtt L_r=I$, which is only the case if $\mathtt E_i=\mathtt E_j$ and $\mathtt L_i = \mathtt L_j$. The former is true if $e_i$ and $e_j$ have the same syndrome $\hat{\vb s}$, while the latter is more complicated to verify, but we know that there are only $2^{2k}$ possibilities in $\mathcal L$ to consider. Ultimately, we may index the degenerate sets based on their syndrome and logical error component, both represented by the tuple $(\hat{\vb s}, \mathtt L)$. We use $e^d_i = \mathtt E_i \mathtt L_i$ to refer to the actual representative of $e_i$ (and its degenerate equivalents) during the correction process, since we know that if we can correct $e^d_i$ by applying the unitary $(e^d_i)^\dagger$ to the circuit, so can we indirectly also correct $e_i$, since $(e^d_i)^\dagger e_i = \mathtt S_i$, and stabilizers act as the identity on the encoded quantum state.

\begin{figure}[t]
    \includegraphics[width=\linewidth]{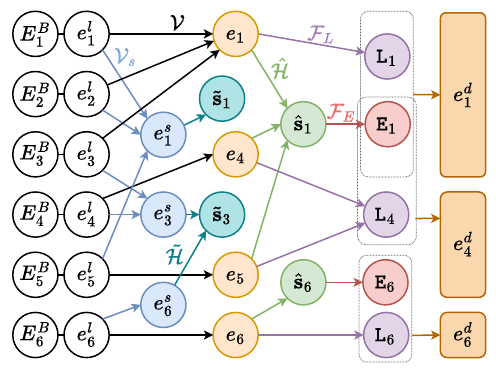}
    \caption{Relations between the different quantities of interest. For simplicity, every error represented is assumed to be a base error.
    }
    \label{fig:error_relations}
\end{figure}

For an error $E_i$ arising from syndrome extraction $g$, there are two associated syndromes of interest, instead of one. The syndrome
\begin{equation}
    \hat{\vb s}_i = \vb e_i \vb A^T\label{eq:hs}
\end{equation}
corresponds to the syndrome obtained from a subsequent noiseless syndrome extraction after extraction $g$, that is, after the error has occurred. In this case, we can consider the error model to be similar to the one used in \cite{Cruz_Access_2023}, where the (propagated) error pattern $e_i$ is present before any stabilizer is applied for the syndrome extraction process. Instead of using \cref{eq:hs}, we may alternatively compute $\hs_i$ by first computing the ancilla error pattern. If we think of $e_i$ as a different local error pattern $e^l_j$, then, following \cref{eq:error_propagation}, we must have
\begin{equation}
    Ve_i V^\dagger = Ve_j^l V^\dagger = e^s_j \otimes e_i,\label{eq:ts}
\end{equation}
and $\hs_i$ is given by the $X$ component (i.e. second half in $[Z|X]$ format, see \cref{eq:binary_repr}) of $\vb e_j^s$. This type of syndrome is always zero for preparation and measurement errors, since measurement errors do not affect subsequent extractions. Under this latter formalism, we may observe this by noting that if $e_i=I$, then necessarily $e^s_j = I$ and $\hs_i = \vb 0$.  

Beyond this typical syndrome, we also have the syndrome $\ts_i$ obtained from the same extraction $g$ where the error $E_i^B$ occurred. For instance, if the error occurs at the last implemented CNOT gate, it is likely that $\ts_i=0$. Unlike $\hs_i$, this syndrome is non-zero for simple measurement errors. In general, this syndrome contains less information than $\hs_i$, since, for errors later in the extraction, many of the syndrome bits will be zero, as the stabilizers were applied before the error occurred. Following the second approach previously presented to compute $\hs_i$ (see \cref{eq:ts}), we have that $\ts_i$ is given by the $X$ component of the original ancilla qubit pattern $\vb e_i^s$ (in binary form).

Both $\ts$ and $\hs$ refer to a $(n-k)$ bit string, corresponding to the syndrome that could be obtained from a single syndrome extraction. As both these syndromes can be deterministically obtained from the error $E$ of interest, we use the simple notation $\ms$ to specifically refer to the syndrome that is measured during the syndrome extraction. As previously stated, for errors stemming from only one extraction (labeled $g$), we have either $\ms=\ts$ (if the measured syndrome comes from the noisy extraction), $\ms=\vb 0$ (if it comes from a previous extraction) or $\ms=\hs$ (if from a subsequent extraction).

In general, compound errors may stem from multiple syndrome extractions. We use superscript notation to indicate the extraction index, in order to distinguish it from the error index (which is a subscript). When there are $q$ syndrome extractions, we refer to the total list of measured syndromes by $s:=\qty{\vb s^1, \ldots, \vb s^q}$. If some base error $E^{B, g}_i$ occurs at extraction $g$, we expect to measure the syndrome sequence given by
\begin{equation}
s_i = \{\ldots, \vb 0^{g-1}, \ts_i^g, \hs_i^{g+1}, \ldots, \hs_i^{q}\}.    
\end{equation}
Note that, for compound errors, the syndrome sequences of the constituting errors may be combined. If $E_r = E_iE_j$, then $s_r = s_i \oplus s_j$, where the modular sum operation is applied element-wise, to all $q$ syndromes. Then, for $E_j = E_{i_1}^B \cdots E_{i_\omega}^B$, we have
\begin{equation}
    \mathcal Q(E_j) := s_j = s_{i_1} \oplus \cdots \oplus s_{i_\omega}.\label{eq:Q}
\end{equation}
In particular, if errors $E_i$ and $E_j$ occur at extractions $g$ and $h$ ($h>g$), respectively, and no other errors occur, then we would expect to measure the syndrome sequence
\begin{multline}
s = \{\ldots, \vb 0^{g-1}, \ts_i^g, \hs_i^{g+1}, \ldots, \\
\hs_i^{h-1}, \hs_i^h \oplus \ts_j^h, \hs_i^{h+1} \oplus \hs_j^{h+1}, \ldots\}.    
\end{multline}
As a result, we observe that $\ts_i, \ts_j, \hs_i, \hs_j$ do not directly provide the full information necessary to identify the compound error $E_i^g E_j^h$ that occurred. In the general case where there are $q$ syndrome extractions, we may require all $q$ measured syndromes to optimally correct errors.

For compound errors $E_r$ stemming from multiple syndrome extractions, the syndrome $\ts_r$ is undefined, but $\hs_r$ may still be defined as
\begin{equation}
    \hs_r = \bigoplus_{j = 1}^\omega \hs_{i_j},
\end{equation}
where $\hs_{i_j}$ are the syndromes of the constituting base errors. Similarly, $e_r = e_{i_1}\cdots e_{i_\omega}$. The decomposition in \cref{eq:error_decomposition} and subsequent analysis is also applicable.

The notation is summarized in \cref{fig:error_relations}, with an example given in \cref{fig:model_relations}.

In \cref{fig:error_relations}, the mappings $\mathcal V_s$, $\mathcal V$, $\hat{\mathcal H}$, $\tilde{\mathcal H}$, $\mathcal F_E$ and $\mathcal F_L$ are mostly independent, and generally non-injective. $\mathcal V$ and $\mathcal V_s$ stem from \cref{eq:error_propagation}, and provide the main and ancilla error patterns $e_i$ and $e_i^s$. Concretely, we have $e_i = \mathcal V (e^l_i) = (Ce^l_iC^\dagger)_m$ and $e_i^s = \mathcal V_s (e^l_i) = (Ce^l_iC^\dagger)_a$, where $m$ and $a$ stand for the main and ancilla subspaces, respectively. Using \cref{eq:ts} yields $\tilde{\mathcal H}$ and $\hat{\mathcal H}$. The latter can also be implemented by using \cref{eq:hs}. Concretely, we have $\hs_i = \hat{\mathcal H}(e_i) = \text{bin}(e_i)\vb A^T = \text{comp}_X(\text{bin}((Ve_ iV^\dagger)_a))$ and $\ts_i = \tilde{\mathcal H}(e_i^s) = \text{comp}_X(\text{bin}(e^s_i))$ (see \cref{eq:bin,eq:comp}). The $\hat{\vb s}_i$ and $\vb s_i$ obtained by the syndrome extraction process can then be used to try and determine $e_i^d$, the representative pattern of the degenerate set to which $e_i$ belongs. By applying $e_i^d$ to the noisy quantum state, we correct the effect of the error $E_i$. The mappings $\mathcal F_E$ and $\mathcal F_L$ are quite involved, so their description is delegated to \cref{sec:decoding}.

\section{Decoding}\label{sec:decoding}
The error pattern statistics given by the noise model of \cref{sec:setup} lead to a very high number of degenerate error patterns (see \cref{sec:degeneracy_example} for examples). As a result, the approximation made in \cite{Cruz_Access_2023}, where codes were approximated to be non-degenerate, is no longer accurate, as it would significantly underestimate the code's capabilities. In this section, we modify the decoding procedure in \cite{Cruz_Access_2023} to account for error degeneracy. The modified procedure is optimal in principle. It has previously been shown that such optimal procedures must be \#P-complete in general \cite{iyerHardnessDecodingQuantum2015,hsieh,kuo}. Since we are applying the decoding procedure to random codes with no exploitable structure, our decoding procedure has poor scaling capabilities for high entropy noise and large code sizes. Nonetheless, following \cite{Cruz_Access_2023}, we hope to show it to be of interest in regimes of small code size or low entropy, so it is still worth exploring the decoding properties of this optimal procedure. It is also possible (though not covered in this work) for the decoding complexity to be greatly improved with simpler approximations and heuristics to the optimal approach. The optimal decoding procedure is summarily presented in \cref{alg:optimal_decoding}.

\begin{algorithm}[H]
\caption{Error processing}
\label{alg:data}
\begin{algorithmic}[1]
\REQUIRE $\mathcal{N}$
\ENSURE A data table $D$
\STATE Initialize empty data table $D$
\FORALL{$(p_i, E_i) \in \mathcal{N}$}
    \STATE Compute $e_i$, $\hs_i$, and $s_i$
    \STATE Compute $\mathtt{L}_i$ associated with $e_i$
    \STATE Compute $e^d_i$
    \IF{$e^d_i$ not in any entry in $D[s_i]$}
        \STATE Store $(e^d_i, p_i)$ in $D[s_i]$
    \ELSE
        \STATE Add $p_i$ to $p$ in entry $(e^d_i, p)$
    \ENDIF
\ENDFOR
\end{algorithmic}
\end{algorithm}

\begin{algorithm}[H]
\caption{Optimal decoding}
\label{alg:optimal_decoding}
\begin{algorithmic}[1]
\REQUIRE $\mathcal{N}$
\ENSURE A decoding table $T$
\STATE Initialize empty decoding table $T$
\STATE $D \leftarrow \textsc{Data}(\mathcal{N})$
\FORALL{entry $s$ in $D$}
    \STATE Set $T[s]$ as the pattern $e^d$ with highest $p$ in $D[s]$
\ENDFOR
\end{algorithmic}
\end{algorithm}

When considering this optimal decoding procedure, we note that, while we focus on the noise model in \cref{fig:faulty_noise_model}, the decoding procedure is naturally applicable to models where there are additional sources of error independent of the syndrome extractions themselves. For that case, the noise statistics would simply include those additional errors.

Moreover, the decoding procedure presented in this section does incorporate any assumptions about the underlying nature of the noise, as it meant to be a fully general procedure. In particular, we do not wish to assume that higher order errors are less likely than lower order ones, as there may be practical regimes where particular high order errors dominate the noise statistics (such as burst errors). In subsection \cref{sec:Bernoulli_noise_model}, we adapt the general decoding procedure to the particular noise model described in \cref{sec:setup}.

For the decoding, we require a procedure that, given a syndrome sequence $s$, outputs the error pattern $e^d$ that needs to be applied to the circuit to correct the most likely source of error. Since we are interested in the analysis of the decoding procedure in the optimal case, and less concerned about practical limitations, we assume that such a procedure corresponds to a decoding table $T$, storing the $s, e^d$ pairs.

The decoding table $T$ may be obtained as follows (see \cref{alg:data,alg:optimal_decoding} for pseudo-code, and \cref{fig:decoding_example} for an example):
\begin{enumerate}
    \item For each error $E_i$ (with corresponding probability $p_i$), we compute its syndrome sequence $s_i$, and also $e_i$ and $\hs_i$ (corresponding to the mapping $\mathcal Q$ in \cref{eq:Q}, and the mappings $\mathcal V$ and $\hat{\mathcal H}\circ \mathcal V$, respectively, in \cref{fig:error_relations}, for base errors). Since the circuit is a stabilizer circuit, it can be efficiently simulated \cite{aaronson_improved_2004}, and these quantities efficiently computed. 
    \item We compute $\mathtt E_i = \mathcal F_E(\hs_i)$ and $\mathtt L_i = \mathcal F_L(e_i)$. As we already know $\hs_i$, we end up with the degenerate set $(\hs_i, \mathtt L_i)$ (and its representative $e_i^d=\mathtt E_i \mathtt L_i$), to which $e_i$ belongs.
    \item We repeat steps 1 and 2 for all errors in $\mathcal N$. Starting with an empty data table $D$, for each error $E_i$, we add $(e_i^d, p_i)$ to the entry $D(s_i)$. If an entry with $e^d_i$ already exists, we add $p_i$ to the entry's probability. This procedure results in the data table $D$.
    \item For each syndrome sequence $s=\qty{\ms^1, \ldots, \ms^q}$ in $D$, we choose the degenerate set with the highest associated probability as the actual coset leader, that is, the one that is corrected if $s$ is measured.
    \item The resulting syndrome table $T$ then acts as our decoding method.
\end{enumerate}
This decoding is optimal because, for any given syndrome, there is no other way for the decoding to be more successful than as described here, since this method already picks the most probable degenerate set $(\hs, \mathtt L)$, given the only information available \emph{a priori}, which is $\mathcal N$ and the observed syndrome $\vb s$. It is optimal under the reasoning that we consider any unsuccessful correction to be a complete failure, with no possible partial success.

While the procedure as described is done in series, iteratively traversing the $E_i$ errors, it can be trivially parallel, by splitting the error list across multiple parallel workers. See \cref{sec:parallel_decoding} for a full description. The pseudo-code for the parallel implementation can be seen in \cref{alg:parallel_decoding}.

To implement the decoding procedure (in particular step 2), \emph{a priori}, we require the efficient implementation of two functions:
\begin{itemize}
    \item A function $\mathcal F_E: \hat{\vb s} \mapsto \mathtt{E}_{\hat{\vb s}}$ that, for a given code, and taking a syndrome $\hs$ as input, outputs a deterministic error pattern $\mathtt{E}_{\hat{\vb s}}$ that can act as a coset leader for the syndrome $\hs$. That is, any error pattern $e_i$ with syndrome $\hs$ can be decomposed (following \cref{eq:error_decomposition}) using the error component $\mathtt{E}_{\hat{\vb s}}$. Having access to this function considerably reduces the required serialized processing for the decoding, and the required memory, as we don't need to keep track of tentative coset leaders as we iterate through the errors $E_i$, and we can be sure that different parallel workers have the coset leader in the same degenerate set (in fact, we can be sure that they are equal).
    \item A function $\mathcal F_L: e_i \mapsto \mathtt L_i$ that, for a given code, and taking an error pattern $e_i$ as input, deterministically outputs the logical component $\mathtt L_i$ of the degenerate set to which this error pattern belongs to.
\end{itemize}

These functions are described in detail in \cref{sec:function_F_E,sec:function_F_L}. 

\subsection{Function \texorpdfstring{$\mathcal F_E$}{FE}}\label{sec:function_F_E}
When analyzing the code, instead of working with the $(n-k)$ minimal stabilizers $\qty{S_i}$ we extracted from the encoding $U$ (see \cref{eq:default_stabilizers}), we work with a different set $\qty{S_{i,\rm rre}}$. Each stabilizer in this new set can be thought of as some combination of the stabilizers in $\qty{S_i}$. To be more specific, considering that $S_i$ corresponds to row $i$ of the quantum check matrix $\vb A$ (with size $(n-k)\times 2n$), $\qty{S_{i,\rm rre}}$ corresponds to row $i$ of $\vb A$ in reduced row echelon form (also known as canonical form). We can convert $\vb A$ to reduced row echelon form because products of stabilizers are still stabilizers. Since we are in $\mathbb F_2$, adding or subtracting rows of $\vb A$ is equivalent to multiplying stabilizers. As long as the resulting matrix is full rank (which is always the case, since the procedure for converting to reduced row echelon form preserves rank), the resulting new matrix $\Arre$ encodes a new set of minimal stabilizers, $\qty{S_{i,\rm rre}}$, in its rows.

In practice, we can imagine that the measured syndrome $\vb s$ gets converted to the ``reduced row echelon'' syndrome $\vb s_{\rm rre}$, which can be done with a $(n-k)\times (n-k)$ matrix that encodes the steps needed to convert $\vb A$ to reduced row echelon form. Let this matrix be $\vb J$. We have $\Arre = \vb J\vb A$ and $\vb s_{\rm rre} = \vb J \vb s$.

Working with $\Arre$, let $h_i$ be the index of the pivot of row $i$ (it's not guaranteed that $h_i=i$, since the pivots may not all be along the main diagonal of $\Arre$). Since $\qty{S_{i,\rm rre}}$ comes from reduced row echelon form, the stabilizer $S_{i,\rm rre}$ will be the only minimal stabilizer with a nonzero entry at index $h_i$. Then, if $1\leq h_i \leq n$, the error $Z_{h_i}$ necessarily yields the syndrome bit 1 for $S_{i,\rm rre}$ and zero for all other minimal stabilizers in $\qty{S_{i,\rm rre}}$. If $n+1\leq h_i \leq 2n$, the error $X_{h_i-n}$ necessarily yields the syndrome bit 1 for $S_{i,\rm rre}$ and zero for all other minimal stabilizers. Consequently, we can use these $(n-k)$ errors as a basis to construct a deterministic error pattern $\mathtt E_{\hs}$ for every syndrome $\hs$. Let $\mathtt E_i$ be the error associated with $S_{i,\rm rre}$ ($\mathtt E_i$ equals $Z_{h_i}$ or $X_{h_i-n}$, as described). Since the code is linear, for any syndrome $\vb s_{\rm rre}$, if $i_1,\ldots, i_k$ are the indices where the syndrome $\vb s_{\rm rre}$ is 1, then the compound error $e=\mathtt E_{i_1}\ldots \mathtt E_{i_k}$ must necessarily have syndrome $\vb s_{\rm rre}$. In other words, $\vb s_{\rm rre}= \vb e \Arre^T$.

Since $e=\mathtt E_{\vb s_{\rm rre}}$ when using the minimal stabilizers $\qty{S_{i,\rm rre}}$, and $\vb J$ is a linear transformation, then we must have $e=\mathtt E_{\hs}$ when using the minimal stabilizers $\qty{S_{i}}$. Therefore, if the code has stabilizers $\qty{S_i}$, then $\mathcal F_E$ implements the procedure
\begin{equation}
    \hat{\vb s} \implies \vb s_{\rm rre} \implies e = \mathtt E_{\hat{\vb s}}.
\end{equation}
Because of the construction of $e$, for all error patterns with the same syndrome, the same error component is computed. The error $\mathtt E_{\hat{\vb s}}$ then acts as the error component for all error patterns with syndrome $\hs$, for the decoding.

\subsection{Function \texorpdfstring{$\mathcal F_L$}{FL}}\label{sec:function_F_L}

\begin{algorithm}[H]
\caption{Finding logical component}
\label{alg:logical_component}
\begin{algorithmic}[1]
\REQUIRE $\bm{e_i}', \bm{A}_{\text{rre}}, \bm{L}_{\text{rre}}$
\ENSURE Vectors $\bm{u_A}, \bm{u_L}, \bm{v'}$
\STATE Initialize $\bm{u_A}, \bm{u_L}$ to zero vectors
\STATE Let $\bm{v} \leftarrow \bm{e_i}'$
\STATE Let $H_A$ (resp. $H_L$) be an ordered list of pivot positions of $\bm{A}_{\text{rre}}$ (resp. $\bm{L}_{\text{rre}}$)
\FORALL{$h_i$ in $H_A$}
    \IF{entry $h_i$ of $\bm{v}$ is 1}
        \STATE Subtract row $i$ of $\bm{A}_{\text{rre}}$ from $\bm{v}$ (mod 2)
        \STATE $[\bm{u_A}]_i \leftarrow 1$
    \ENDIF
\ENDFOR
\STATE Let $\bm{v'} \leftarrow \bm{v}$
\FORALL{$l_i$ in $H_L$}
    \IF{entry $l_i$ of $\bm{v}$ is 1}
        \STATE Subtract row $i$ of $\bm{L}_{\text{rre}}$ from $\bm{v}$ (mod 2)
        \STATE $[\bm{u_L}]_i \leftarrow 1$
    \ENDIF
\ENDFOR
\STATE \textbf{return} $\bm{u_A}, \bm{u_L}$
\STATE (if the computation was correct, then $\bm{v}$ should equal $\bm{0}$)
\end{algorithmic}
\end{algorithm}

For the function $\mathcal F_L$, we need to consider the different degenerate sets. We know that each error $e_i$ can be decomposed in terms of the coset leader $\mathtt E_i$, some stabilizer $\mathtt S_i$, and some logical operator $\mathtt L_i$ (see \cref{eq:error_decomposition}). There are $4^k$ logical operators (including the identity) and each identifies one of the $4^k$ degenerate sets associated with each syndrome $\hs$.

Consequently, we can create a one-to-one mapping between the logical operators and the degenerate sets. Since we can determine the tentative coset leader $\mathtt E_i$ with $\mathcal F_E$, we only need to determine the logical component $\mathtt L_i$ that composes our input error pattern.

With this in mind, we continue the approach from \cref{sec:function_F_E}. We use the row echelon form of $\vb A$, that is, $\Arre$, so we work with $\qty{S_{i,\rm rre}}$ instead. We perform the same procedure to the logical operators. Let $\vb{\bar X}_i$ and $\vb{\bar Z}_i$ be the binary row vector representations of $\bar X_i$ and $\bar Z_i$, respectively (see \cref{eq:default_logical_X,eq:default_logical_Z}), in $[X|Z]$ format. Let $\vb L$ be the $2k \times 2n$ binary matrix that encodes the original $2k$ minimal logical operators (see \cref{eq:default_logical_X,eq:default_logical_Z}), that act as generators to the $4^k$ total logical operators. Row $i$ of $\vb L$ is given either by $\vb{\bar X}_i$, if $1\leq i \leq k$, or by $\vb{\bar Z}_{i-k}$, if $k+1\leq i \leq 2k$.

Just as with the stabilizers, we know that products of logical operators are also logical operators. Moreover, products of a logical operator with stabilizers correspond to the same logical operator. Since we are working in $\mathbb F_2$, adding or subtracting rows of $\Arre$ or $\vb L$ is equivalent to multiplying stabilizers and operators in Pauli string form. Considering the augmented $(n+k)\times 2n$ matrix $\mqty[\Arre \\ \vb L]$, by using the rows of $\Arre$ and $\vb L$, we may put the $\vb L$ component in its row echelon form, $\vb L_{\rm re}$. We can then put $\vb L_{\rm re}$ in reduced row echelon form using only the rows of $\vb L_{\rm re}$, yielding $\vb L_{\rm rre}$. Note that we cannot use the rows in $\vb L_{\rm re}$ to further simplify $\Arre$, as the resulting rows would no longer correspond to stabilizers. The procedure may be represented as
\begin{align}
    \mqty[\Arre \\ \vb L_{\rm rre}] &= \mqty[I_{n-k} & 0 \\ 0 & \vb J_L] \mqty[\Arre \\ \vb L_{\rm re}]\\
    &= \mqty[I_{n-k} & 0 \\ 0 & \vb J_L] \mqty[I_{n-k} & 0 \\ \vb J_A & I_{2k}] \mqty[\Arre \\ \vb L],
\end{align}
with $I_p$ a $p\times p$ identity matrix. The matrices $\vb J_A$ and $\vb J_L$ are of size $2k\times (n-k)$ and $2k\times 2k$, respectively, and just like $\vb J$ in \cref{sec:function_F_E}, they represent the linear transformation required to put the matrix in reduced row echelon form.

The resulting rows of $\vb L_{\rm rre}$ correspond to $2k$ (possibly different) generators for the logical operators. These generators may no longer satisfy the anti-commutation relations expected of $\bar X$ and $\bar Z$, but they are not required to. The final $\vb L_{\rm rre}$ matrix is such that the columns with the same index as the pivots of $\Arre$ are zero, and the columns with the pivots of $\vb L_{\rm rre}$ have only one non-zero element, its pivot.

For every error pattern $e_i$ with syndrome $\vb s_i$, we know that its error component $\mathtt E_i$ (given by $\mathcal F_E$) is such that $e_i\mathtt E_i = \mathtt S_i \mathtt L_i=: e_i'$, for some unknown $\mathtt S_i$ and $\mathtt L_i$. Consequently, to determine the degenerate set to which $e_i$ belongs, we only need to decompose $e_i'$ into its $\mathtt S_i$ and $\mathtt L_i$ components. Concretely, we are looking for the unique row vectors $\vb u_A$ (of size $n-k$) and $\vb u_L$ (of size $2k$) such that
\begin{align}
    \vb e_{i,[X|Z]}' &= \mqty[\vb u_A & \vb u_L] \mqty[\Arre \\ \vb L_{\rm rre}].
\end{align}
where $\vb e_i'$ is exceptionally in $[X|Z]$ format. 
Since $\Arre$ and $\vb L_{\rm rre}$ are already in reduced row echelon form, finding the two vectors is straightforward. The procedure is described in \cref{alg:logical_component}. Once the $\vb u_A$ and $\vb u_L$ row vectors are determined, the $\mathtt S_i$ and $\mathtt L_i$ components are simply given by (see \cref{eq:bin})
\begin{align}
    \mathtt S_i &= \text{op}(\vb u_A \Arre)\\
    \mathtt L_i &= \text{op}(\vb u_L \vb L_{\rm rre}).
\end{align}

Alternatively, we may simply skip the computation of $\vb u_L$ in \cref{alg:logical_component}. Let $\vb v'$ equal the computed row vector $\vb v$ just after $\vb u_A$ is computed, but before the iteration through the pivots of $\vb L_{\rm rre}$. Then, we equivalently have $\mathtt L_i = \text{op}(\vb v')$.

The full procedure
\begin{equation}
    e_i \implies e_i\mathtt E_i \implies \vb u_L\; (\text{or }\vb v') \implies \mathtt L_i
\end{equation}
corresponds to the function $\mathcal F_L$.

\section{Asymptotic regime}\label{sec:asymptotic}

We can estimate the optimal performance we can obtain from the decoding procedure by looking at how it performs as the number of extractions considered is increased. We are interested in computing the limit where we have infinite extractions, where the decoding would be optimal. Although this regime is impossible to attain in practice, we expect that, as we increase the number of extractions, the decoding dynamics should converge to the asymptotic corresponding to that optimal case, allowing us to estimate the code's performance in that regime.

We consider the total probability of correction failure $P_{\rm total}$ to correspond to the probability that an error is not completely corrected. That is, the correction chosen does not correspond to the right degenerate set. Note that this definition provides a lower bound on the fidelity $F$ of the resulting quantum state, given by
\begin{equation}
    F \geq 1 - P_{\rm total},
\end{equation}
since it effective treats any unsuccessful correction as producing a state with zero fidelity, whereas in practice the uncorrected error may not produce an orthogonal quantum state. Nonetheless, it is a useful lower bound often used in the literature \cite{fowler_surface_2012, Cruz_Access_2023}, and that we choose to use here as well.

\begin{figure}[t]
    \includegraphics[width=\linewidth]{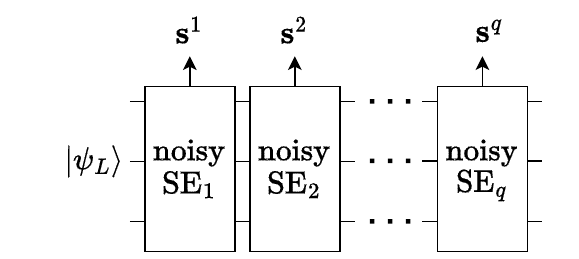}
    \caption{Model considered to compute the asymptotic regime, where $q\rightarrow \infty$.}
    \label{fig:fit_model}
\end{figure}

Let $P_\infty$ be the asymptotic limit of $P_{\rm total}$ when the number of syndrome extractions $q$ goes to infinity (see \cref{fig:fit_model}). Since there are $L=2^{2k}$ degenerate sets associated to each syndrome sequence $s$, then, regardless of the encoding used, for a given $s$, the probability $p$ of an error having occurred that is in the most likely degenerate set satisfies $p \geq 1/L$. 
If $e^c_i$ is the error pattern representative of the most likely degenerate set for $s_i$, then we must have
\begin{align}
    P_\infty &= 1 - \sum_i p(e^c_i e_i \in \mathcal S|E_i)p(E_i)\\
    &\leq 1 - \sum_i (1/L)p(E_i)\\
    &= 1 - 1/L\\
    &= 1 - 2^{-2k},
\end{align}
with equality achieved for the case of maximum noise entropy.

When comparing two setups, $A$ and $B$, with $q_A$ and $q_B=q_A+h$ $(h>0)$ syndrome extractions, respectively, we necessarily have
\begin{equation}
    P_{\rm total}(A) \leq P_{\rm total}(B)
\end{equation}
since introducing $h$ additional noisy syndromes extractions introduces additional errors in the model, which may or may not be correctable.

We are interested in determining the failure probability $P_{\rm failure}$ induced by a single additional syndrome extraction, preceded and succeeded by an arbitrarily high number of extractions. In the context of fault-tolerance, we wish to determine if, for a given $p$, $P_{\rm failure}$ increases or decreases with increasing qubit count $n$. We expect, for low (resp. high) values of $p$, $P_{\rm failure}$ decreases (resp. increases) with $n$, with a phase transition at some $p_{\rm threshold}$, to be determined.

Since we do not have direct access to $P_{\rm failure}$, it must be computed from the measured value of $P_{\rm total}$. We develop an effective model to quantitatively relate the two quantities.

Consider a variant of the $B$ setup above, labeled $B'$, where the first $q_A$ extractions are solely used to identify and correct errors stemming from implementing those extractions, and similarly, the last $h$ extractions are solely used to deal with errors resulting from the $h$ extractions, for a total of $q_B=q_A+h$, as before. In other words, in the $B'$ setup, the procedure is partitioned into to separate and independent decoding processes.

We expect
\begin{equation}
    P_{\rm total}(B) \leq P_{\rm total}(B').
\end{equation}
since, in the $B$ setup, the last $h$ extractions also provide information about errors in the previous extractions, which can lead to a more successful decoding. The $B'$ setup does not use this information. 

In fact, if an error occurs at extraction $q_A$, we expect that a lower number $h$ of subsequent extractions will result in higher failure rates (per extraction), since the decoding process has less information to correctly identify the error. Since we are interested in the limiting case where the number of subsequent extractions is infinite (for any given extraction where an error can occur), we wish to discount the effect of limited syndrome information from the calculated probability of failure $P_{\rm failure}$. We label the errors that could have been successfully corrected with $h\rightarrow\infty$ but were not with low $h$ \emph{escaped errors}.

For low $q$, we expect (see \cite{dennis_topological_2002}) that escaped errors (in particular errors with $\ts = 0$ but $\hat{\vb s}\neq 0$) can be identified with additional syndromes from more subsequent extractions, so that each new extraction reduces the number of escaped errors by a factor $\delta$. Of course, new extractions also introduce new escaped errors, and errors at the end extractions are more likely to escape correction.

Consider a setup with $q$ extractions as a setup with $q-1$ extractions preceded by one additional extraction. The errors of this additional extraction will be detected by the $q-1$ subsequent extractions, so that only a factor $\delta^{q-1}$ escape through the whole setup incorrectly identified. Therefore, applying the reasoning recursively for $q$ extractions, we expect the probability of an error passing through the extractions undetected to be given by
\begin{align}
    P_u(p,q) &\simeq P_u(p,q-1) + \delta^{q-1}P_u(p,1) \\
    &\simeq P_u(p,1) \frac{1-\delta^q}{1-\delta}. \label{eq:P_u_q}
\end{align}

\begin{figure}[t]
    \includegraphics[width=\linewidth]{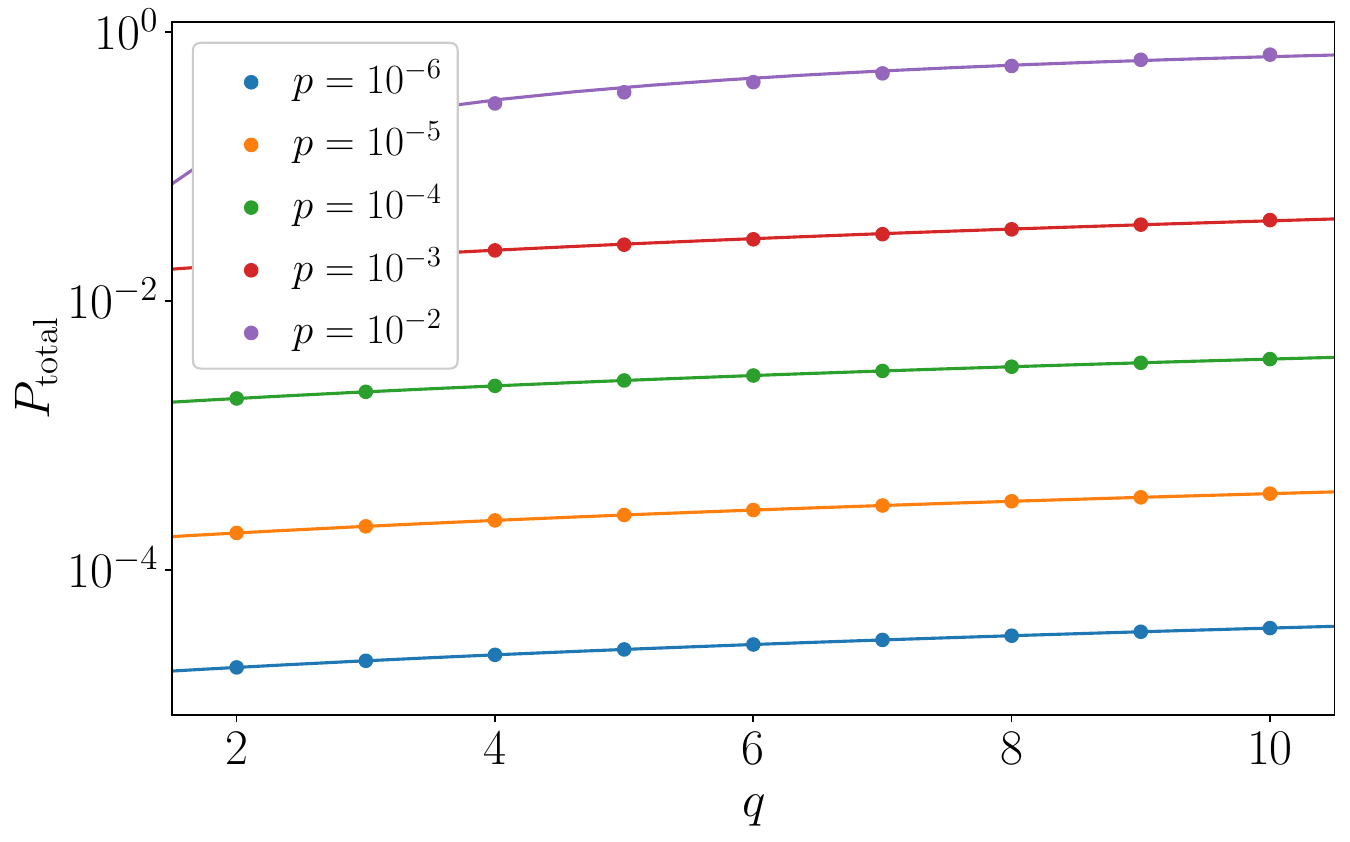}
    \caption{Experimental $P_{\rm total}$, and resulting fit by \cref{eq:fit}. We observe the experimental data matching the qualitative description of the theoretical analysis, with $R^2>0.999$ for all cases.}
    \label{fig:fit_test}
\end{figure}

For $q\gg 1$, we expect that the probability of successful correction with $q$ extractions $P_{\rm succ}(p,q)$ 
will be approximately the probability of success with $q-1$ extractions times the per-extraction probability of success (which accounts for the additional extraction). We have
\begin{align}
     P_{\rm succ}(p,q) &\simeq P_{\rm succ}(p,q-1)(1-P_{\rm failure}(p)) \\
     &= (1-P_{\rm failure}(p))^q \\
     \text{with }&P_{\rm succ}(p,0)=1.
\end{align}
However, since $P_{\rm failure}$ does not incorporate escaped errors, the true probability of success $P_{\rm succ}'$ is given by
\begin{equation}
    P_{\rm succ}'(p,q) = P_{\rm succ}(p,q)(1-P_u(p,q)),
\end{equation}
yielding
\begin{equation}
    P_{\rm total}(p,q) \simeq 1-(1-P_{\rm failure}(p))^q(1-P_u(p,q)).
\end{equation}
If $\delta\ll 1$, which is expected, then, when using $q\gg 1$ total extractions, we may approximate $P_u(p,q)\simeq P_u(p,\infty)$, leading to
\begin{align}
    P_{\rm total}(\p,q) &\simeq 1-C(\p)(1-P_{\rm failure}(\p))^q,\label{eq:fit}\\
    \text{with }C(\p) &:= 1-P_u(p,\infty),\label{eq:C_p}
\end{align}
where $C$ incorporates the escaped errors. See \cref{fig:fit_test} for a numerical verification of this model. Once more, note that we discount these errors, and their probability $P_u$, from the failure probability $P_{\rm failure}$, since we are interested in the per-extraction failure probability, and $P_u$ constitutes a global effect. As previously indicated, if a syndrome extraction is followed by $h\rightarrow \infty$ extractions, then the probability that an error stemming from that extraction goes completely undetected is $\delta^h P_u(p, 1) \rightarrow 0$. 

$P_{\rm failure}$ is the asymptotic contribution of each syndrome extraction to the probability of failure. We have that, by fitting, for each $\p$, 
\begin{gather}
    Y = m X + b,\\
    \text{with }Y := \log(1-P_{\rm total}(p,q)),\quad X := q,
\end{gather}
then, from \cref{eq:fit},
\begin{equation}
    P_{\rm failure}(p) = 1- e^m,\quad C(p) = e^b,\label{eq:P_fail_C}
\end{equation}
for each $\p$ considered.

\begin{figure}[t]
    \includegraphics[width=\linewidth]{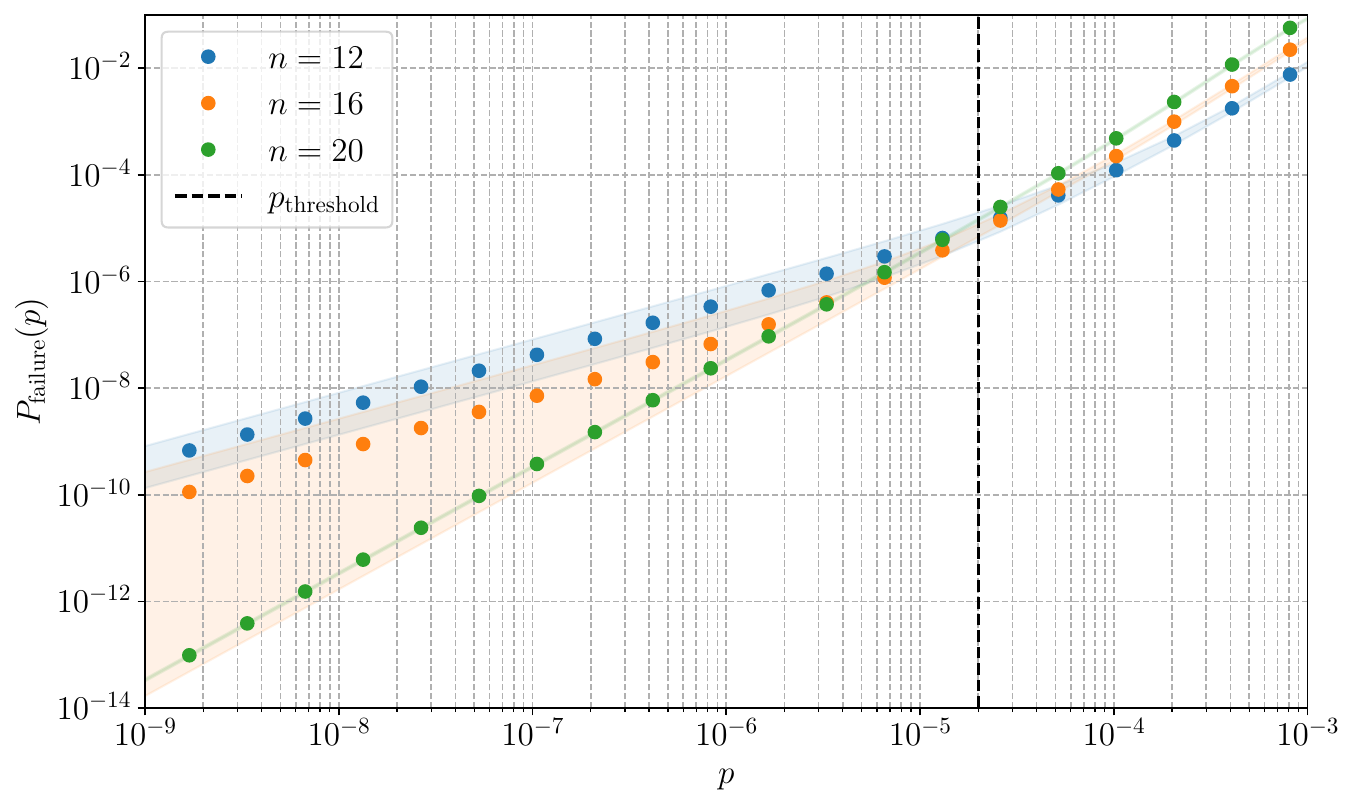}
    \caption{Performance of QRLCs with $k=1$, using fault-tolerant QGRAND. We observe that the asymptotic version presents a threshold around $\p \simeq \pnum$. The shaded regions denote a 50\% confidence interval (from 50 samples), and the dots are the median performance observed. The vertical line indicates the $\pth$ value.}
    \label{fig:ft_p_threshold_QRLC}
\end{figure}

See \cref{fig:ft_p_threshold_QRLC} for results. We observe $\pth \simeq \pnum$, suggesting the QGRAND technique can be used in the fault-tolerant regime. Nonetheless, we note that this optimal decoding procedure is not scalable for high entropy noise models, such as those given by $\p \gtrsim \pth$. Therefore, to use these techniques, we are forced to either simplify our model, turning the decoding procedure suboptimal, or to focus on a regime where it can be applied in practice.

For reference, we also estimate the performance of the uncoded case. Considering the number of noisy CNOT gates present in a circuit with $q$ iterations to be $\N(q)$, we get
\begin{equation}
    P_{\rm failure, uncoded}(\p) \simeq 1 - (1-\p)^{\N(1)}.
\end{equation}
$\N$ doesn't include preparation and measurement errors, as these don't propagate throughout the circuit.

The simulation is performed for $\omega < \omega_{\rm max}$, and approximated for the higher $\omega$ values. For high $\omega$, we use the reasonably accurate assumption that error patterns are uniformly assigned to the syndrome sequences. Therefore, before computing $P_{\rm failure}$, we modify $P_{\rm total}$ with the correction
\begin{equation}
    P_{\rm total}' = P_{\rm total} - \frac{P(\omega>\omega_{\rm max})}{2^{n+k}},
\end{equation}
where $2^{n+k}$ is the number of degenerate sets associated with each syndrome sequence. The correction is negligible for high $n$, but may play a noticeable role for $\p \simeq 1$.

\section{Discussion and conclusions}\label{sec:conclusion}

In this work, we extend the decoding procedure for QRLCs introduced in \cite{Cruz_Access_2023} to explicitly account for error degeneracy. Consequently, our technique constitutes a maximum likelihood decoding procedure, which is guaranteed to be optimal. We analyze the fault-tolerant characteristics of QRLCs with the presented decoding technique, by accounting for preparation, measurement, and gate errors in the syndrome extraction procedure itself, and observe a $\pth \simeq \pnum$ in the asymptotic limit. As far as the authors are aware, this is the first proposed decoding technique in the literature for quantum random linear codes in the fault-tolerant regime, where preparation, measurement, and gate errors are not considered negligible.

We note that this decoding procedure is not equivalent to finding the lowest weight error pattern associated with each syndrome, as might be done by more standard algorithms, since a faulty CNOT gate error effect can propagate considerably through the circuit before being possible to detect it, so that, by the time it is detected, its error pattern is no longer low weight.

In this work, we have removed the channel errors present in \cite{Cruz_Access_2023} and considered only the main error sources associated with the syndrome extraction process. In particular, we considered preparation and measurement errors in the ancilla qubits, and two-qubit gate errors. Although this is a common approach to take when studying the fault-tolerance capabilities of different codes \cite{fowler_surface_2012}, it leads to unrealistic results for high code rates. In the limit when $R\rightarrow 1$, the syndrome extraction process has a negligible number of minimal stabilizers, and as a result negligible error sources, under this noise model. Consequently, in this regime, higher code rates lead to lower $P_{\rm failure}$, not because of better correction capabilities, but because error sources decrease faster than the correction capabilities do, as the code rate increases.

Moreover, we have analyzed the asymptotic regime of infinite syndrome extractions. Although impractical, these asymptotic results enable us to study the behavior of the optimal decoding procedure, as previously described. Nonetheless, practical limitations might impose suboptimal steps in the decoding process, and obviously a finite number of syndrome extractions.

To account for these limitations, we must consider that, in practice, there are non-trivial computing steps performed between syndrome extractions (such as logical gates for quantum computing, and Bell-pair creation for quantum communication) that introduce their own errors independently from the syndrome extraction steps. When accounting for this additional error source, we expect the pathological behavior for high code rates to disappear. In future work, we intend to study these more practical regimes. Furthermore, we assumed that all-to-all connectivity (between any of the $n$ qubits) is possible in practice. This assumption is required for the scaling results in \cite{brown_short_2013}, and is used in this work. Nonetheless, it may be dropped for practical reasons, as the more recent results in \cite{Gullans_Krastanov_Huse_Jiang_Flammia_2021} suggest.

As previously mentioned, the noise guessing decoding procedure is expected to be viable only in situations of low noise entropy and low $n$. Even disregarding the limitations imposed by the asymptotically large number of syndrome extraction, it is also the case that the noise entropy increases rapidly as $n\rightarrow \infty$ and $k=1$, as considered for the fault-tolerance analysis. This is a known limitation of the decoding procedure. For this reason, and for the fact that better known codes, such as surface codes, have higher $\pth$ values, we do not expect the decoding procedure described in this work to be competitive in those regimes. Although it remains to be confirmed in future work, we conjecture that, given the optimal decoding properties of the described procedure, it may be worthwhile to employ in scenarios where code versatility is needed, the noise statistics are not approximately fixed, and the code rate is desired to be very high. In those cases, we expect the method to have similar use cases to those previously described in \cite{Cruz_Access_2023}, as the additional gates used by the syndrome extraction process would not have a strong impact on decoding performance.

Beyond the straightforward approach described in \cref{alg:optimal_decoding}, we may also wish to sacrifice the decoding optimality for the sake of decoding throughput, or lower hardware requirements, rendering the decoding process more easily scalable. This can be achieved with either known techniques, such as compressive sensing or deep learning, or with more straightforward approaches, such as greedy variants of the method. For instance, for the Bernoulli noise model in this work, if we take the coset leader to be the first error pattern associated with a syndrome, we will end up with a suboptimal version of \cref{alg:optimal_decoding}, equivalent to \cite{Cruz_Access_2023}, but one which reduces the memory requirements by as much as a factor of $4^k$. There are also specific simplifications that can be used to implement the decoding procedure faster when the noise model has some exploitable structure, as is the case with the noise model considered here. We plan to cover some of these approaches in future work.

\appendices
\crefalias{section}{appendix}
\crefalias{subsection}{appendix}

\section{Reducing stabilizer weight}\label{sec:reducing_stabilizer_weight}

Although we are working with quantum random linear codes, which have little exploitable structure \emph{a priori}, we note that the minimal stabilizers can be efficiently chosen to have weight lower than the average of $3n/4$. To do so, we may take the original minimal stabilizer arising from the technique described in \cite{Cruz_Access_2023}, represent them with the parity check matrix, and put the matrix in canonical form, which is equivalent to reduced row echelon form. The new simpler minimal stabilizers correspond to the rows of the resulting matrix. 

If the pivots of the matrix in reduced row echelon form are all in the first $n$ columns, then this technique reduces the weight of the non-$Z$ components of each minimal stabilizer to at most $1+k$, and $1+k/2$ on average. If $k$ is low and $n$ is large, this technique can result in a considerable reduction in the weight of the minimal stabilizers, as their average weight goes from $3n/4$ to $3k/4 + (n-k-1)/2 + 1=(2n+k+2)/4$. Instead of each term being equally distributed between $I,X,Y,$ and $Z$ as before, here only the indices greater than $n-k$ maintain that distribution, and we have, for $S_i$, index $i$ equally distributed between $X$ and $Y$, and index $j\leq n-k, j\neq i$ equally distributed between $I$ and $Z$. 

This structure simplifies the application of the syndrome extraction process, as it reduces the number of CNOT gates from $\sim 3n(n-k)/4$ to $(2n+k+2)(n-k)/4$, and similarly reduces the number of 1-qubit gates. Despite these benefits, in our numerical analysis we have not assumed such an approach was taken in the circuit implementation, in order not to introduce unwarranted structure in the noise statistics, as we are interested in analyzing the more general scenario. 

Nonetheless, as explained in \cref{sec:function_F_E,sec:function_F_L}, we have used this simplification in our decoding implementation, when given the noise statistics associated with the unsimplified stabilizers. As they generate the same stabilizer group, they are mathematically equivalent for the same given noise statistics, and we expect either approach to lead to similar numerical results.

\section{Degeneracy analysis}\label{sec:degeneracy_example}

Regarding degeneracy, following the notation introduced in \cref{sec:notation}, we consider errors $E_i$ and $E_j$ to be degenerate if and only if $e_ie_j \in \mathcal S$. Nonetheless, we may describe three types of degenerate errors, all prevalent in the noise model considered in this work.

\paragraph{Identical errors.}
These are errors such that $e_i=e_j$ and $e_i^s=e_j^s$. For example, since we are considering the model implementation where the CNOT gates of the conditional stabilizer are implemented in succession, with the control always being the ancilla qubit (see \cref{fig:faulty_noise_model}), then any error of the form $Z_c I_t$ (with $c$ and $t$ the control and target qubit indices, respectively) will commute with subsequent CNOT gates controlled by the qubit $c$. Therefore, this component does not add any error terms to the main $n$ qubits, and instead simply negates the measured ancilla qubit. As a result, for any error term without this component, there is an error term with this component where the error pattern in the main $n$ qubits is the same, and the ancilla qubit is simplify negated. Given this degeneracy, the problem reduces to two scenarios: one where there is an even number of such errors, where the syndrome is unaffected, and one with an odd number of such errors, where the ancilla bit is negated. Among these two classes, all errors are not only degenerate, but identical. 

\paragraph{Pseudo-identical errors.}
Besides the identical errors, we also observe cases where $e_i=e_j$ but $e_i^s\neq e_j^s$ (with $\text{comp}_X(e_i^se_j^s)=\vb 0$, otherwise the errors would have different syndromes).

\paragraph{Non-identical errors.}
We also have the more general case where $e_i\neq e_j$ (with either $e_i^s= e_j^s$ or $e_i^s\neq e_j^s$), while still retaining $e_ie_j \in \mathcal S$.

\begin{figure}[t]
    \includegraphics[width=\linewidth]{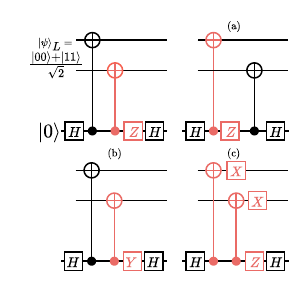}
    \caption{For the example code in \cref{fig:model_relations}, we may observe the three forms of degenerate errors. Considering $E_i$ as the error showcased in the top left circuit, the identical, pseudo-identical, and non-identical errors can be seen in subfigures (a), (b), and (c), respectively.}
    \label{fig:degeneracy_example}
\end{figure}

See \cref{fig:degeneracy_example} for an example. There, the code's sole non-trivial stabilizer is $X_1X_2$. We have $e_i^s=e_a^s=a_c^s=X$, $e_b^s=Y$ (ignoring the phase), $e_i=e_a=a_b=I_1I_2$, and $e_c=X_1X_2$.

\section{Analytical threshold without degeneracy}\label{sec:analytical_threshold}

For this analysis, we disregard preparation and measurement errors, as the derivation can be readily extended to the complete model.

Keeping $k$ constant, the value of $n$ will determine the number of CNOT gates ($\N$) in the circuit, and the code's correction capabilities. We may estimate its performance by considering the approximation given by random ideal codes, as in \cite{Cruz_Access_2023}.

For $S=:2^{n-k}\gg 1$ and $N$ the number of distinct errors, the equation
\begin{equation}
    f = \frac{S}{N+1}\qty[1-\qty(1-\frac{1}{S})^{N+1}]\label{eq:f_def}
\end{equation}
may be approximated by
\begin{align}
    f &\simeq \frac{1-e^{-r}}{r}, \qquad\qquad\qquad\qquad r := \frac{N+1}{S}\\
    \implies f &\simeq \frac{1-(1-r+r^2/2)}{r}=1-\frac{r}{2}\text{ for }r \ll 1,
\end{align}
since, from \cref{eq:f_def}, we have
\begin{align}
    f &=\frac{1}{r}\qty[1-\qty[\qty(1-\frac{1}{S})^S]^{r}]\\
    &\simeq\frac{1}{r}\qty[1-(e^{-1})^r].
\end{align}
Since $S=2^{n-k}$, this indicates that
\begin{align}
    f &\geq 1-\epsilon\\
    \iff N &\lesssim 2^{n-k+1}\epsilon.\label{eq:N_err}
\end{align}

W.l.o.g., consider that each CNOT gate can only suffer from a specific error, instead of 15. For the Bernoulli noise model we are considering, the error order $\omega$ is given by the binomial distribution
\begin{equation}
    \omega \sim \mathcal B(\N,\p)
\end{equation}
For $\p$ fixed, and as $\N \rightarrow \infty$, this distribution can be approximated by
\begin{equation}
    \mathcal N(\N\p,\N\p(1-\p))
\end{equation}
using the De Moivre-Laplace theorem.

Suppose we start by correcting the lowest order errors (which are more likely to occur), and we wish to correct the errors up to order $\omega_{\rm max}$ such that we have the probability $(1-\epsilon)$ of correcting an error we observe. For a normal distribution, this is given by the quantile function
\begin{equation}
    Q(p) = \mu + \sigma \sqrt{2} \text{erf}^{-1}(2p-1).
\end{equation}
The error function erf cannot be easily approximately. Nonetheless, we may observe that it can approximated by
\begin{equation}
    \text{erf}(x) \simeq 1-\alpha\exp(-(x-\beta)^2),\text{ for }x\gg 1,
\end{equation}
leading to
\begin{equation}
    \text{erf}^{-1}(x) \simeq \beta +\sqrt{\log\qty(\frac{\alpha}{1-x})}\text{ for }1-x\ll 1.
\end{equation}
For simplicity, we consider $\alpha=1, \beta=0$, which does not meaningfully affect our conclusions here. The quantile function then becomes
\begin{align}
    Q(1-\epsilon) &\simeq \mu + \sigma \sqrt{2} \qty(\beta +\sqrt{\log\qty(\frac{\alpha}{2\epsilon})})\\
    &\simeq \mu + \sigma \sqrt{2\log\qty(\frac{1}{2\epsilon})}\label{eq:tmax_quantile}.
\end{align}
with
\begin{align}
    \mu &=\N\p\\
    \sigma &= \sqrt{\N\p(1-\p)}\\
    &\simeq \sqrt{\N\p}.
\end{align}
Now that we have $\omega_{\rm max} = Q(1-\epsilon)$, we need to estimate the number of errors $\tilde N$ up to order $\omega_{\rm max}$, given by
\begin{equation}
    \tilde N = \sum_{j=0}^{\omega_{\rm max}} \mqty(\N \\ j).
\end{equation}
Unfortunately, there is no closed form expression for this value. However, for $\omega_{\rm max} \ll \N$, this is roughly equal to
\begin{align}
    \tilde N &\simeq \mqty(\N \\ \omega_{\rm max})\\
    &\simeq \frac{\N^\N}{{\omega_{\rm max}}^{\omega_{\rm max}}(\N-\omega_{\rm max})^{\N-\omega_{\rm max}}}\notag\\
    &\phantom{\qquad\qquad}\times\sqrt{\frac{\N}{2\pi\omega_{\rm max}(\N-\omega_{\rm max})}}.
\end{align}
This may be simplified down to
\begin{equation}
    \tilde N = 2^{\N h_2(\omega_{\rm max}/\N)} \sqrt{\frac{1}{2\pi\omega_{\rm max}}},\label{eq:tilde_N}
\end{equation}
where $h_2$ is the Shannon entropy.

From \cref{eq:N_err,eq:tilde_N}, we therefore conclude that
\begin{align}
    \log N &\sim \tilde O(n)\\
    \log \tilde N &\sim \tilde O(\N),
\end{align}
where $\tilde O$ denotes Big-O notation up to log factors. 
Since $N$ indicates the code's correction capabilities, while $\tilde N$ indicates the necessary number of errors that the code needs to correct to preserve $P_{\rm failure}$, then we must have $N \gtrsim \tilde N$ and consequently $\N \sim \order{n}$. This is verified for some common codes, such as surface codes, but for our implementation we have
\begin{equation}
    \N = \frac{3}{4}n(n-k) \sim \order{n^2},
\end{equation}
so we conclude that, if all errors are non-degenerate, the code does not have a visible $p_{\rm threshold}$, since an increase in $n$ increases $P_{\rm failure}$, regardless of $\p$.

However, we actually observe a threshold for QRLCs. This is thanks to the fact that the noise statistics given by the noise model of \cref{sec:setup} actually lead to a very high number of degenerate errors. Therefore, in practice, the number of distinct errors grows with $\order{n}$ and not $\order{n^2}$ as indicated by the analysis above. We later confirm this scaling for escaped errors, in \cref{sec:P_derivation}.

With this insight in mind, we modify the QGRAND algorithm in \cite{Cruz_Access_2023} to account for the possibility of degenerate errors. The modified algorithm is presented in \cref{sec:decoding}.

\section{Applying error correction}\label{sect:applying_error_correction}

Certain QECCs, such as surface codes, are designed so that a physical error correction is actually unnecessary to implement, as all changes can be made in software, classically \cite{fowler_surface_2012}. If the whole quantum circuit is unitary, then this procedure can actually be implemented in general: instead of correcting the error, we leave the affected state as is and simply XOR any subsequent syndrome $\vb s$ with the identified error syndrome $\vb e$, that is, $\vb s \mapsto \vb s \oplus \vb e$. As a result of this, we only need to keep track of these detected errors classically, in order to correct the subsequent syndromes.

In any case, since the correction portion is always single-qubit gates, we assume that their contribution to the total error is negligible, so we can disregard this trick for now, for the sake of simplicity. As a result, the procedure to apply the error correction is the same as in \cite{Cruz_Access_2023}.

\section{Parallel decoding}\label{sec:parallel_decoding}

The decoding procedure described in \cref{sec:decoding} can be performed in parallel. If there are $W$ parallel workers available, we start by splitting the entries in $\mathcal N$ into $W$ parts of equal size, labeling each $\mathcal N_w, 1\leq w\leq W$. Each set $\mathcal N_w$ is then processed independently by an individual procedure according to the procedure described in the \cref{sec:decoding}, thereby yielding the data table $D_w$.

All the $W$ data tables $D_w$ may then be merged to generate the full data table $D$, from which the decoding table $T$ can be straightforwardly computed. See \cref{alg:parallel_decoding} for a description of the parallel decoding procedure.

% Algorithm 1: Parallel decoding
\begin{algorithm}[H]
\caption{Parallel decoding}
\label{alg:parallel_decoding}
\begin{algorithmic}[1]
\REQUIRE $\mathcal{N}, W$
\ENSURE A decoding table $T$
\STATE Initialize empty data table $D$ and decoding table $T$
\STATE Split $\mathcal{N}$ into $W$ sets (almost) equal in size, labeled $\mathcal{N}_w$ ($1 \leq w \leq W$)
\FORALL{$w$ parallel workers}
    \STATE Initialize empty data table $D_w$
    \STATE $D_w \leftarrow \textsc{Data}(\mathcal{N}_w)$
\ENDFOR
\STATE $D \leftarrow \bigcup_w D_w$
\FORALL{entry $s$ in $D$}
    \STATE Set $T[s]$ as the pattern $e^d$ with highest $p$ in $D[s]$
\ENDFOR
\STATE \textbf{return} $T$
\end{algorithmic}
\end{algorithm}

\section{Full decoding example}\label{sec:full_example}

\begin{figure}[t]
    \includegraphics[width=\linewidth]{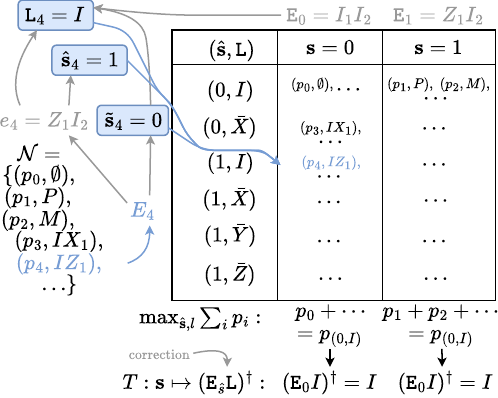}
    \caption{Optimal decoding of the code in \cref{fig:model_relations} (but with only one syndrome extraction), as described in \cref{alg:optimal_decoding}.}
    \label{fig:decoding_example}
\end{figure}

For the example in \cref{fig:decoding_example}, we consider the encoding gate to be $U=\text{CNOT}(2,1)H_2$, with the starting qubit at index 1. $P$ and $M$ stand for preparation and measurement error in the only ancilla qubit $a$, respectively. $AB_j)$ is the error corresponding to ``error $A_aB_j$ occurred at CNOT gate $j$''. Following \cref{alg:optimal_decoding}, we iterate through the errors in $\mathcal N$ and, using the functions $\mathcal F_E$ and $\mathcal F_L$, determine where to put them in the table $D$. Once we have iterated through all errors, we compute the optimal $(\hs, \mathtt L)$ entry, or alternatively, $e^d$ representative, and delete the remaining entries, yielding the decoding table $T$. 

Given $U$, and following \cref{eq:default_stabilizers,eq:default_logical_X,eq:default_logical_Z}, the minimal stabilizer is $S_1 = X_1X_2$, and the logical operators are $\bar X = I_1 X_2$ and $\bar Z = Z_1Z_2$. This choice of encoding leads to minimal stabilizers and logical operators such that the augmented matrix is
\begin{equation}
    \mqty[\vb{A}\\ \vb{L}] = \mqty[\vb{A}\\ \vb{\bar{X}}\\ \vb{\bar{Z}}] = \mqty[1 & 1 & 0 & 0 \\ 0 & 1 & 0 & 0 \\ 0 & 0 & 1 & 1] = \mqty[\Arre \\ \vb{L}_{\rm rre}],
\end{equation}
since, in this simple example, we already have $\vb A = \Arre$ and $\vb L = \vb{L}_{\rm rre}$, so $\vb J = I$. Here, $\vb{A}, \vb{\bar{X}},$ and $\vb{\bar{Z}}$ are the binary representations of the minimal stabilizers and logical operators, in $\mqty[X|Z]$ format.

Following the procedure in \cref{sec:function_F_E}, we have
\begin{equation}
    \mathtt E_0 = I_1I_2 \qquad \mathtt E_1 = Z_1I_2.
\end{equation}

Let's consider the full procedure in \cref{sec:decoding} applied to the specific error in \cref{fig:model_relations}, such as
\begin{equation}
    E = ``\text{Error }I_a Z_1\text{ in CNOT gate }1\text{''.}
\end{equation}
Associated to this error, we have the quantities
\begin{align}
    \ts &= 0\\
    e &= Z_1 I_2\\
    \vb e &= \mqty[0 & 0 & 1 & 0]\\
    \hs &= \vb e \vb A^T = 1\\
    \hs_{\rm rre} &= \hs = 1\\
    e \mathtt E_{\hs} &= I_1 I_2 = \mathtt S \mathtt L\\
    \implies \mathtt L &= I_1 I_2 = I.
\end{align}

Applying the procedure in \cref{sec:decoding} and \cref{alg:optimal_decoding}, we obtain the tentative decoding table in \cref{tab:decoding_table_example}. In the table, for the sake of simplicity, we represent preparation and measurement errors by $P$ and $M$, respectively. $AB_i$ represents the CNOT gate error where error $A_a B_i$ occurs just after the noiseless CNOT gate. For simplicity, compound error are not shown.

\begin{table}[H]
\centering
\caption{Decoding for the code in \cref{fig:model_relations}.}
\label{tab:decoding_table_example}
\setlength{\tabcolsep}{1pt}
\begin{tabular}{@{}l|c|cc@{}}
\textbf{$(\hat{\mathbf{s}}_{\rm rre},\mathtt L)$} & $e^d$ & \textbf{$\mathbf{s}=0$}  & \textbf{$\mathbf{s}=1$}        \\ \hline
$(0, I)$              &  $I_1I_2$        & $\emptyset, XI_2, XX_1$  & $P,\!M,\!ZI_2,\!ZI_1,\!YX_1,\!YI_2$ \\
$(0,\bar X)$          &  $I_1X_2$        & $IX_1,\!XI_1,\!IX_2,\!XX_2$ & $ZX_1, YI_1, ZX_2, YX_2$       \\
$(0,\bar Y)$          &  $Z_1Y_2$        &                          &                                \\
$(0,\bar Z)$          &  $Z_1Z_2$        &                          &                                \\
$(1, I)$              &  $Z_1I_2$        & $IZ_1, XY_1$             & $ZZ_1,YY_1$                    \\
$(1,\bar X)$          &  $Z_1X_2$        & $XZ_1, IY_1$             & $YZ_1, ZY_1$                   \\
$(1,\bar Y)$          &  $Z_1Y_2$        & $IY_2, XY_2$             & $ZY_2, YY_2$                   \\
$(1,\bar Z)$          &  $I_1Z_2$        & $IZ_2, XZ_2$             & $ZZ_2, YZ_2$                  
\end{tabular}
\end{table}

The preparation and measurement errors have probability $\p$ of occurring, and the gate errors have probability $\p/15$. If we assume that it is very unlikely that no error has occurred (i.e. $\p$ is high), then the final decoding table (when only considering errors of order 1) is given by \cref{tab:final_decoding_table_example}. In the table, if $\ms = 0$ is measured, the most likely degenerate set to have occurred is $(0, \bar X)$, with probability $4\p/15$ (see \cref{tab:decoding_table_example}). We assume that $\p$ is such that the no-error case is unlikely, otherwise $(0, \bar I)$ would be the most likely degenerate set. If $\ms = 1$ is measured, the most likely degenerate set to have occurred is $(0, \bar I)$, with probability $34\p/15$. For each case, the error pattern that should be applied to the quantum state to correct the error is given by $e^d= \mathtt E_{\hs} \mathtt L$.

\begin{table}[H]
\centering
\caption{Final decoding table for \cref{fig:model_relations}.}
\label{tab:final_decoding_table_example}
\begin{tabular}{c|c}
\textbf{$\mathbf{s}=0$}  & \textbf{$\mathbf{s}=1$}        \\ \hline
$(0, \bar X)$  & $(0, I)$ \\
$\downarrow$ & $\downarrow$ \\
$\mathtt E_0 \bar X = I_1 X_2$ & $\mathtt E_0 I = I_1 I_2$
\end{tabular}
\end{table}

\section{Simplified noise statistics}\label{sec:simplified_statistics}

When considering a Bernoulli noise model, such as in \cref{sec:Bernoulli_noise_model}, including preparation and measurement errors into the noise model breaks some of the structure of the noise statistics, since not all base errors will be equally likely. It also makes the decoding process harder to simulate. In the simpler setup where only gate errors occur, if we have $g$ CNOT gates affected with an error, with each error having probability $p_{\rm CNOT}/15$ of occurring, then the number of errors and their individual probability would be given by
\begin{align}
    N_E(g) &= 15^g \mqty(N_{\rm CNOT} \\ g)\\
    P(g) &= \qty(\frac{p_{\rm CNOT}}{15})^{g} (1-p_{\rm CNOT})^{N_{\rm CNOT} - g}.
\end{align}
These formulas stem from the fact that each CNOT gate has 15 associated errors, and only one of these may occur at a time. For preparation and measurement errors, there is only one error pattern per ancilla qubit: either there is a bit flip, or there isn't. If there are $A$ ancilla qubits (generally, $A=n-k$), and a preparation or measurement error occurs with probability $p_M$, these same quantities are given by
\begin{align}
    N_E(\omega) &= \sum_{g=0}^\omega 15^{g} \mqty(N_{\rm CNOT} \\ g) \mqty(2A \\ \omega-g)\label{eq:N_E_t}\\
    P(\omega,g) &= \qty(\frac{p_{\rm CNOT}}{15})^{g} (1-p_{\rm CNOT})^{N_{\rm CNOT} - g}\label{eq:P_tj}\\
    &\qquad\qquad \times p_M^{\omega-g}(1-p_M)^{2A-\omega+g},
\end{align}
where $\omega$ indicates the total error order, and $g$ indicates the error order when ignoring preparation and measurement errors.
Note that these expressions are somewhat more complicated. In particular, for errors of order $\omega$, we now need to keep track of the distinct number of CNOT ($g$) and preparation/measurement ($\omega-g$) base errors that occurred, instead of just one parameter.

Fortunately, if we are using optimal decoding (that properly accounts for degenerate errors), there is a quick-and-dirty way to mimic the simpler noise statistics associated with only having gate errors. If we use $p_M = p_{\rm CNOT}=:p$ (which is a relatively common choice in the literature, see \cite{fowler_surface_2012}, and the one used in \cref{sec:setup}), then we can consider that each preparation and measurement error is a CNOT gate error. Instead of there being only one error per qubit (corresponding to the possible bit flip), we consider that there are 15, all equal in nature, and each occurring with probability $p_M/15$. These cloned errors will be degenerate among themselves, so the optimal decoding procedure will analyze this setup correctly. The total number of error patterns $N_E(\omega)$ will be overcounted, but we never actually use it directly for the decoding procedure, so the overcounting does not constitute an issue. In this formulation, we may pretend that we have no preparation and measurement errors, and that we have instead
\begin{equation}
    \Tilde{N}_{\rm CNOT} := N_{\rm CNOT} + 2A\label{eq:tilde_N_CNOT}
\end{equation}
CNOT gates in the syndrome extraction circuit. 
The probability associated with each error will be correct, yielding
\begin{equation}
    P(\omega) = \qty(\frac{\p}{15})^\omega (1-\p)^{\tilde{N}_{\rm CNOT} - \omega}.\label{eq:P_omega}
\end{equation}

\section{Bernoulli noise model improvements}\label{sec:Bernoulli_noise_model}

\begin{algorithm}[H]
\caption{Data table $\tilde{D}_1$ for Bernoulli noise}
\label{alg:bernoulli_data_1}
\begin{algorithmic}[1]
\REQUIRE $\mathcal{B}$
\ENSURE A data table $\tilde{D}_1$
\STATE Initialize empty data table $\tilde{D}_1$
\STATE $\bm{n_0} \leftarrow (1, 0)$
\STATE Store $\{I: \bm{n_0}\}$ in $\tilde{D}_1[\{\bm{0}, \ldots, \bm{0}\}]$
\FORALL{$E_i^B$ in $\mathcal{B}$}
    \STATE Compute $e_i$, $\hat{s}_i$, and $s_i$
    \STATE Compute $\mathcal{L}_i$ associated with $e_i$
    \STATE Compute $e^d_i$
    \IF{$e^d_i$ not in $\tilde{D}_1[s_i]$}
        \STATE $\bm{n_i} \leftarrow (0, 0)$
        \STATE $\bm{n_i}(1) \leftarrow 1$
        \STATE Store $\{e^d_i: \bm{n_i}\}$ in $\tilde{D}_1[s_i]$
    \ELSE
        \STATE $\bm{n_i}(1) \leftarrow \bm{n_i}(1) + 1$
        \STATE Update $\{e^d_i: \bm{n_i}\}$ in $\tilde{D}_1[s_i]$
    \ENDIF
\ENDFOR
\STATE \textbf{return} $\tilde{D}_1$
\end{algorithmic}
\end{algorithm}

For the special case where each CNOT gate has the same probability $p$ of suffering an error, as described in \cref{sec:setup}, the decoding procedure can be made much more efficient. This procedure may be adapted to other Bernoulli-like noise statistics, but here we focus on this error model.
We can optimize this decoding process in order to avoid having to iterate through all compound errors. Since the technique relies on the inherent structure of errors with the same probability, here we employ the reformulation detailed in \cref{sec:simplified_statistics} to treat preparation and measurement errors as additional gate errors.

The procedure for the list of base errors (with $\omega=1$) is similar to the one described in the beginning of \cref{sec:decoding}. Instead of using the full noise statistics $\mathcal N$, instead we consider a list of base errors
\begin{equation}
    \mathcal B := \qty{E^B_1, E^B_2, \ldots}.
\end{equation}
Given the Bernoulli noise model, all base errors have the same associated probability of occurring, given by $p/15$, so it doesn't need to be stored in $\mathcal B$ (we exclude the no error case). Once we have the data table $D_1$ for $\omega=1$, we can start to optimize the analysis for compound errors.
Instead of iterating through compound errors individually, we iterate through the degenerate sets obtained in the $\omega=1$ step. We also preserve the probability associated with observing an error pattern from each degenerate set in the table. For $\omega=1$, and given the reformulation of \cref{sec:simplified_statistics}, all errors $\hat E_i^B$ may be considered to have a probability $P(1)$ of occurring (see \cref{eq:P_omega}), so the probability associated with each degenerate set in the data table is given by 
\begin{equation}
    p_i := N_i P(1),
\end{equation}
where $N_i$ is the number of errors $E_i$ that can be corrected by applying the coset leader $e_i^d$ associated with the degenerate set of $\hat E_i^B$.   
In summary, we may restructure the data table $D_1$ obtained with \cref{alg:optimal_decoding} (before the final degenerate set selection)
\begin{multline}
    D_1 = \{s_1:\qty{(e^d_{11}, p_{11}), (e^d_{12}, p_{12}),\ldots},\\ s_2:\qty{(e^d_{21}, p_{21}), (e^d_{22}, p_{22}),\ldots}, \ldots\}
\end{multline}
to encode the count $N_i$ instead of $p_i$, and to store a list of error counts for different orders, yielding
\begin{multline}
    \tilde D_1 = \{s_1:\qty{e^d_{11}:\vb n_{11}, e^d_{12}:\vb n_{12},\ldots},\\ s_2:\qty{e^d_{21}:\vb n_{21}, \ldots}, \ldots\}\label{eq:decoding_table_omega_1}
\end{multline}
with
\begin{equation}
    \vb n_i := (0,N_{i},0,\ldots)
\end{equation}
a list of size $(\omega+1)$, where only the second entry of the list (corresponding to $\omega=1$) starts with non-zero entries. The first entry is only non-zero for the $E_0=I$ case, when $\vb s= \hs = 0$ and $\mathtt L=I$, corresponding to the case where no error occurs. We represent the entry of order $g$ by $\vb n_i(g)$. 

Under the formulation of \cref{eq:decoding_table_omega_1}, instead of iterating through the combinations $\qty{E_1^BE_2^B, E_1^BE_3^B,\ldots, E_2^BE_3^B,\ldots}$ as we could do with the naive implementation of \cref{alg:optimal_decoding}, we iterate through the degenerate sets in $\tilde D_1$ as a whole.

Consider the data table $\tilde D_{\omega-1}$ that includes the errors up to order $\omega-1$. To obtain the table for errors up to order $\omega$, we iterate through the combination of the degenerate sets in $\tilde D_{\omega-1}$ and $\tilde D_1$. If the noise statistics are highly degenerate (which is generally the case following the noise model in \cref{sec:setup}), we can have considerable computational savings, since we only need to perform $|\tilde D_{\omega-1}||\tilde D_1|$ computations instead of $15^{\omega}\mqty(N_{\rm CNOT}+2(n-k) \\ \omega)$ (see \cref{sec:simplified_statistics}). While we expect the latter to grow quickly with $\order{n^{2\omega}}$, the former approach should grow, at worst, with $\order{n^\omega}$, and it may grow more slowly in practice.

With this approach we generally overcount the number of errors $N_i$ associated with each degenerate set. There are three types of overcounting:
\begin{itemize}
    \item Counting permuted copies: Consider an order-$(\omega-1)$ error $E_{i_1}^BE_{i_2}^B\ldots E_{i_{\omega-1}}^B$ (with $i_1 < i_2 < i_3 < \ldots$), coming from $\tilde D_{\omega-1}$, and the error $\hat E_j^B$, coming from $\tilde D_1$. W.l.o.g., suppose $j < i_1$. Then, for $\tilde D_\omega$, we will not only count the error $E_j^B E_{i_1}^BE_{i_2}^B\ldots E_{i_{\omega-1}}^B$, but also that same error coming from the combination of the errors $E_j^B E_{i_1}^BE_{i_2}^B\ldots E_{i_{\omega-1}}^B \backslash E_{i_k}^B$ and $E_{i_k}^B$. In total, we overcount each order-$\omega$ error $\omega$ times.
    \item Recounting lower order errors: for the error $E_{i_1}^BE_{i_2}^B\ldots E_{i_{\omega-1}}^B$, composing with any $E_{i_k}^B$ ($1\leq k\leq \omega-1$) reduces the error to one of order $\omega-2$, which was previously counted. Each order-$(\omega-2)$ error we counted before will be recounted $\zeta_{\omega-1,1}$ times, where $\zeta$ is given in \cref{eq:zeta}.
    \item Counting two errors occurring in the same CNOT gate: an error of order $\omega$ stems from base errors that occurred on $\omega$ CNOT gates. When composing this error with another, the resulting compound error may have more than one base error occurring at one or more CNOT gates. As this compound error is impossible, it should be discounted.
\end{itemize}

\begin{algorithm}[H]
\caption{Data table $\tilde{D}_{a+b}$}
\label{alg:bernoulli_data_merge}
\begin{algorithmic}[1]
\REQUIRE $\tilde{D}_a, \tilde{D}_b$
\ENSURE A data table $\tilde{D}_{a+b}$
\STATE Initialize empty data table $\tilde{D}_{a+b}$
\FORALL{$(s_i, e^d_i, \bm{n_i})$ in $\tilde{D}_a$}
    \FORALL{$(s_j, e^d_j, \bm{n_j})$ in $\tilde{D}_b$}
        \STATE $\tilde{\bm{n}_{ij}} \leftarrow \bm{n_i} * \bm{n_j}$ \quad \textit{(convolution)}
        \STATE $\bm{n_{ij}} \leftarrow \tilde{\bm{n}_{ij}}$ with overcounting correction
        \IF{$e^d_ie^d_j$ not in $\tilde{D}_{a+b}[s_i \oplus s_j]$}
            \STATE Store $\{e^d_ie^d_j: \tilde{\bm{n}_{ij}}\}$ in $\tilde{D}_{a+b}[s_i \oplus s_j]$
        \ELSE
            \STATE Add $\tilde{\bm{n}_{ij}}$ to vector in $e^d_ie^d_j$ entry in $\tilde{D}_{a+b}[s_i \oplus s_j]$
        \ENDIF
    \ENDFOR
\ENDFOR
\FORALL{$(s_r, e^d_r, \bm{n_r})$ in $\tilde{D}_{a+b}$}
    \STATE $\bm{n_{r}} \leftarrow \tilde{\bm{n_{r}}}$ with overcounting correction
    \STATE Store $\{e^d_r: \bm{n_{r}}\}$ in $\tilde{D}_{a+b}[s_r]$
\ENDFOR
\STATE \textbf{return} $\tilde{D}_{a+b}$
\end{algorithmic}
\end{algorithm}

We can extend this approach further. Instead of constructing the data table in one order increments, if we already have $\tilde D_\omega$, we may combine it with itself to obtain $\tilde D_{2\omega}$, thereby requiring exponentially fewer iterations, as $\omega$ increases, as long as $\omega$ is such that the codes capabilities are not yet saturated, i.e., not all syndromes are assigned to a degenerate set.

In general, if we compute the noise statistics of errors of order $a$ and $b$ to compute those of order $\omega=a+b$, we have
\begin{gather}
    \vb n_{i}(\omega) = \frac{1}{R_{a,b}}\Big[\tilde{\vb n}_i(\omega) - \sum_{\substack{0<k+r\leq b\\ k,r\geq 0}} \zeta_{a,b}(k,r) \vb n_i(c)\Big],\label{eq:bernoulli_count}\\
    \text{with }c := a+b-2k-r,
\end{gather}
where $\tilde{\vb n}_i(\omega)$ is the count obtained for order $\omega$ before the overcounting correction. The auxiliary functions are given by
\begin{align}
    R_{a,b} &:= \mqty(a+b \\ a)\\
    \zeta_{a,b}(k,r) &:= K^k (K-1)^r \xi_{a,b}(k,r)\label{eq:zeta}\\
    \xi_{a,b}(k,r) &:=\mqty(\tN - c\\ k)\mqty(c\\ r, a-k-r, b-k-r),
\end{align}
where $K$ is the number of distinct errors per CNOT gate (in our case, always 15). Note the use of binomial and multinomial coefficients. To incorporate the effect of preparation and measurement errors, we use $\tilde N_{\rm CNOT}=\N+2(n-k)$ and not $\N$, as explained in \cref{sec:simplified_statistics}. See \cref{sec:bernoulli_derivation} for a derivation of these expressions.

The probability associated with the degenerate set with list $\vb n_i$ is given by
\begin{equation}
    p_i = \sum_{j=0}^\omega \vb n_i(j) P(j).\label{eq:p_i}
\end{equation}

Given this procedure to obtain $\tilde D_\omega$, we may obtain the decoding table $T_\omega$ by following \cref{alg:bernoulli_data_1,alg:bernoulli_data_merge,alg:bernoulli_decoding}.

\begin{algorithm}[H]
\caption{Decoding table $T_{a+b}$}
\label{alg:bernoulli_decoding}
\begin{algorithmic}[1]
\REQUIRE $\tilde{D}_{\omega}$
\ENSURE A decoding table $T_{\omega}$
\STATE Initialize empty decoding table $T_{\omega}$
\FORALL{$s$ in $\tilde{D}_{\omega}$}
    \STATE $(j, p) \leftarrow (-1, 0)$
    \FORALL{$(e^d_i, \bm{n_i})$ in $\tilde{D}_{\omega}[s]$}
        \STATE Using $\bm{n_i}$, compute $p_i$ (\cref{eq:p_i})
        \STATE $(j, p) \leftarrow (i, p_i)$ if $p_i > p$
    \ENDFOR
    \STATE Set $T[s]$ as the pattern $e^d_j$, which has the highest $p$ in $D[s]$
\ENDFOR
\STATE \textbf{return} $T_{\omega}$
\end{algorithmic}
\end{algorithm}

\section{Derivation of decoding formulas for Bernoulli noise}\label{sec:bernoulli_derivation}

As stated in \cref{sec:Bernoulli_noise_model}, a straightforward implementation of the procedure described will overcount the number of errors associated to any given syndrome sequence. There are three types of overcounting, which we may analyze separately.

\subsection{Recounting lower order errors}

Suppose we have already computed the correct error count $\vb n_i(j)$ for $0\leq j\leq \omega-1$ (for all degenerate sets), and we are currently trying the determine $\vb n_i(\omega)$.

For the error $E_{i_1}^BE_{i_2}^B\ldots E_{i_{\omega-1}}^B$, composing with any $E_{i_k}^B$ ($1\leq k\leq \omega-1$) reduces the error to one of order $\omega-2$, which was previously counted. To determine how many errors stem from this dynamic, we may note that any fake compound error of order $\omega$ has a corresponding error of order $\omega-2$, which has already been counted in $\vb n_i(\omega-2)$. Similarly, any error $E_i$ counted in $\vb n_i(\omega-2)$ has a corresponding set of fake compound errors that appear in $n_i(\omega)$. As $E_i$ stems from $\omega-2$ base errors, each affecting a different CNOT gate, these fake compound errors must correspond to an error of the form $E_i E_j^B E_j^B$, where $E_j^B$ is a base error from a CNOT gate not present in $E_i$. For each CNOT gate there are $K=15$ associated base errors, so there are a total of
\begin{equation}
    \zeta_{\omega-1,1}(1,0) = K(\tN-(\omega-2))
\end{equation}
fake error compounds that associated with the error $E_i$. Alternatively, defining $\omega = a+k$, with $k=1$, we may also write the total as
\begin{equation}
    \zeta_{a,1}(1,0) = K(\tN-(a-k)).
\end{equation}

The same principle applies to higher order combinations. If a order-$a$ error $E_i$ is composed with a order-$k$ error $E_j$ ($a\geq k$), and the base errors composing $E_j$ all stem from CNOT gates whose same base errors already compose $E_i$, then the resulting compound error will have order $a-k$, instead of $a+k$. The total number of errors is now given by
\begin{align}
    \zeta_{a,k}(k,0) &= \frac{1}{k!}\prod_{j=0}^{k-1} K(\tN-(a-k)-j)\\
    &= K^k \mqty(\tN - (a-k) \\ k).\label{eq:zeta_kb}
\end{align}

Note, however, that, when composing a order-$a$ error $E_i$ with a order-$b$ error $E_j$ ($a\geq b\geq k$), it may be the case that only some, and not all, of the base errors composing $E_j$ appear in $E_i$. In general, there will only be $k$ base errors in common, for all $1\leq k\leq b$. 

In this case, to cover all possible fake errors, we must choose not $k$ CNOT gates out of the $\tN - (a-k)$ gates not related to the order-$(a-k)$ error (as in \cref{eq:zeta_kb}), but instead choose $k$ CNOT gates out of the gates not related to both the order-$(a-k)$ error, but also the $(b-k)$ base errors in $E_j$ that are valid. Therefore, there are a total of $\tN - (a-k) - (b-k)$ CNOT gates from which we must consider $k$ invalid base errors.

Moreover, associated with the order-$(a-k)$ error from $E_i$, there are several possible valid $(b-k)$ base errors stemming from $E_j$. The total number of fake errors is then given by
\begin{align}
    \zeta_{a,b}(k,0) &= K^k \mqty(\tN - (a-k) -(b-k)\\ k)\\
    &\qquad\qquad\qquad\quad\times\mqty((a-k) +(b-k)\\ b-k)\\
    &= K^k \mqty(\tN - c\\ k)\mqty(c\\ b-k),\label{eq:zeta_k}\\
    &\text{with }c := a+b-2k.
\end{align}
Note that, for $b=k$, we have $c=a-k$, so that \cref{eq:zeta_k} trivially reduces to \cref{eq:zeta_kb}. The resulting lower order error will have order $c$. We would need to discount its affect on $\vb n_i(\omega)$ to obtain the correct count. Unfortunately, it would be difficult to determine the original syndromes of the errors that combined to result in the impossible error, as they may have different origins. As an approximation, we use $\vb n_i(c)$ to estimate the error count. The resulting correction is achieved by subtracting $\vb n_i(c)$ times $\zeta$ from $\vb n_i(\omega)$.

\subsection{Counting impossible errors}

If a order-$a$ error $E_i$ is composed with a order-$r$ error $E_j$ ($a\geq r$), and the base errors composing $E_j$ all stem from CNOT gates already associated with the base errors that compose $E_i$, then the resulting compound error will be impossible, since it will contain at least two different base errors associated with the same CNOT gate (one from $E_i$ and one from $E_j$).

For $r=1$, each error in $\vb n_i(a)$ will have
\begin{equation}
    \zeta_{a,1}(0,1) = (K-1)a
\end{equation}
associated impossible order-$(a+1)$ errors, since $E_i$ is composed of $a$ base errors, and for each base error, there are $(K-1)$ different base errors associated to the same CNOT gate.

For a general $r$, we have, for each order-$a$ error
\begin{equation}
    \zeta_{a,r}(0,r) = (K-1)^r \mqty(a\\ r)\label{eq:zeta_rb}
\end{equation}
associated impossible errors.

When composing a order-$a$ error $E_i$ with a order-$b$ error $E_j$ ($a\geq b$), it may be the case that only $r<b$ base errors composing $E_j$ are impossible, with the remaining $(b-r)$ base errors stemming from CNOT gates not related to $E_i$. 

To estimate the number of impossible errors, we may look at $\vb n_i(a+b-r)$. As before, we must choose $r$ CNOT gates out of the $a$ gates related to $E_i$ to count the number of impossible errors. But, as seen in the previous subsection, to must also count the possible $(b-r)$ base errors in $E_j$ that are valid. 
These factors result in
\begin{align}
    \zeta_{a,b}(0,r) &= (K-1)^r \mqty(a\\ r)\mqty(a+b-r\\ b-r)\\
    &= (K-1)^r \mqty(c\\ r, a-r, b-r),\label{eq:zeta_r}\\
    &\text{with }c := a+b-r\\
    &\text{and }\mqty(x+y+z\\ x, y, z):=\frac{(x+y+z)!}{x!y!z!}
\end{align}
the multinomial coefficient. Again, note that \cref{eq:zeta_r} trivially reduces to \cref{eq:zeta_rb} when $b=r$. The resulting lower order error will have order $c$, so, for this case, we would also need to discount its affect on $\vb n_i(\omega)$ by subtracting $\vb n_i(c)$ times $\zeta$.

\subsection{Counting permuted copies}

Consider an order-$(\omega-1)$ error $E_{i_1}^BE_{i_2}^B\ldots E_{i_{\omega-1}}^B$ (with $i_1 < i_2 < i_3 < \ldots$), coming from $\tilde D_{\omega-1}$, and the error $\hat E_j^B$, coming from $\tilde D_1$. W.l.o.g., suppose $j < i_1$. Then, for $\tilde D_\omega$, we will not only count the error $E_j^B E_{i_1}^BE_{i_2}^B\ldots E_{i_{\omega-1}}^B$, but also that same error coming from the combination of the errors $E_j^B E_{i_1}^BE_{i_2}^B\ldots E_{i_{\omega-1}}^B \backslash E_{i_k}^B$ and $E_{i_k}^B$. In total, we overcount each order-$\omega$ error $\omega$ times.

In general, for every order-$\omega$ error, any possible combination of order-$a$ errors and order-$b$ errors that can generate it (with $\omega=a+b$) will appear in the counting. Since there are
\begin{equation}
    R_{a,b}:= \mqty(\omega \\ a) = \mqty(\omega \\ b)
\end{equation}
ways for order-$a$ and order-$b$ errors to generate an order-$\omega$ error, the final counting (after discounting the previous overcounting cases) should be reduced by a factor of $R_{a,b}$.

\subsection{Full expression}

In general, the erroneous errors that the decoding procedure may containing not only repeated base errors ($k>0$), but also base errors stemming from the same CNOT gate ($r>0$). Therefore, these two factors need to be considered together.

Combining the analyses of the previous subsections, we conclude that, when composing a order-$a$ error $E_i$ with a order-$b$ error $E_j$ ($a\geq b$), we can have $k$ base errors $E_j$ already appearing in $E_i$, and $r$ base errors in $E_j$ sharing the same origin CNOT gate as a base error in $E_i$, with $k+r\leq b$.

To count all these errors, we may look that the errors with order $c=(a+b-2k-r)$, from which we can generate all the invalid order-$\omega$ errors. As before, the $k$ repeated base errors are chosen from those associated with CNOT gates that are not related to a valid base error in the compound error. There are $\tN-c$ such gates, and each one has $K$ associated base errors. 

Moreover, the from the $c$ base errors, we may consider that $a-k-r$ (resp. $b-k-r$) correspond to the base errors in $E_i$ (resp. $E_j$) that raise no issue, with the remaining $r$ errors corresponding to base errors stemming from CNOT gates that also originated invalid base errors in $E_j$.

Grouping all three overcounting issues, we end up with
\begin{align}
    \zeta_{a,b}(k,r) = K^k (K-1)^r \xi_{a,b}(k,r)
\end{align}
with
\begin{align}
    \xi_{a,b}(k,r) &:=\mqty(\tN - c\\ k)\mqty(c\\ r, a-k-r, b-k-r),\\
    \text{and }c &:= a+b-2k-r,
\end{align}
which generalizes \cref{eq:zeta_k,eq:zeta_r}.

The corrected count is consequently given by
\begin{gather}
    \vb n_{i}(\omega) = \frac{1}{R_{a,b}}\Big[\tilde{\vb n}_i(\omega) - \sum_{\substack{0<k+r\leq b\\ k,r\geq 0}} \zeta_{a,b}(k,r) \vb n_i(c)\Big],\\
    \text{with }c := a+b-2k-r,
\end{gather}
as indicated in \cref{eq:bernoulli_count}, with $\tilde{\vb n}_i(\omega)$ being the original count from the optimized decoding process. As previously indicated, since the estimate of the impossible errors is not exact, this formula is approximate.

\section{Alternative definitions of \texorpdfstring{$\mathcal F_E$}{FE} and \texorpdfstring{$\mathcal F_L$}{FL}}\label{sec:alternative_definitions}

Instead of using the formulation described in \cref{sec:function_F_E,sec:function_F_L}, we may consider an alternative formulation that, while less computationally efficient, is conceptually more straightforward. Under this formalism, the components $\mathtt E_i$ and $\mathtt L_i$ can be computed at once from $e_i$, so there is less of an need to separate the two processes.

For a given error $E_i$, we compute $e_i$. Taking the encoding $U$, we compute
\begin{equation}
    e_i^u := U^\dagger e_i U.
\end{equation}
The \emph{unencoded error pattern} $e_i^u$ corresponds to the effect of $e_i$ on the quantum state if it were decoded. We may decompose it into
\begin{equation}
    e_i^u := \mathtt E_i^u \mathtt S_i^u \mathtt L_i^u,
\end{equation}
with
\begin{align}
    \mathtt E_i^u &= U^\dagger \mathtt E_i U,\\
    \mathtt S_i^u &= U^\dagger \mathtt S_i U,\\
    \mathtt L_i^u &= U^\dagger \mathtt L_i U.
\end{align}
We may also decompose it into the Pauli string for the first $k$ data qubits ($e_i^D$) and the additional $(n-k)$ redundancy qubits ($e_i^R$),
\begin{equation}
    e_i^u =: e_i^D \otimes e_i^R.
\end{equation}
From \cref{eq:default_stabilizers,eq:default_logical_X,eq:default_logical_Z}, we have that
\begin{align}
    Z_{i+k} &= U^\dagger S_{i} U\\
    X_j &= U^\dagger \bar X_{j} U\\
    Z_j &= U^\dagger \bar Z_{j} U.
\end{align}
Therefore, decoding $e_i$ into $e_i^u$ cleanly separates the different components. $\mathtt L_i^u$ corresponds to the components of $e_i^u$ in the first $k$ qubits. We have
\begin{align}
    \mathtt L_i^u &:= e_i^D \otimes I_{n-k}\\
    &= \qty(\prod_{j=1}^k X_j^{b_j^X})\qty(\prod_{j=1}^k Z_j^{b_j^Z})\\
    \implies \mathtt L_i &= U (e_i^D \otimes I_{n-k}) U^\dagger\\
    &= \qty(\prod_{j=1}^k {\bar X}_j^{b_j^X})\qty(\prod_{j=1}^k {\bar Z}_j^{b_j^Z}),
\end{align}
where $b_j^X$ and $b_j^Z$ are the $X$ and $Z$ components of $\vb e_i^D$, respectively.

Let $e_i^{R,X}$ and $e_i^{R,Z}$ be the $X$ and $Z$ components of $e_i^R$, respectively (where the $Y$ components have been decomposed into $X$ and $Z$, as in \cref{eq:pauli_string}), so that $e_i^R = e_i^{R,X}e_i^{R,Z}$ (disregarding the phase factor). We have
\begin{align}
    \mathtt E_i^u &:= I_{k} \otimes e_i^{R,X}\\
    &= \prod_{i=k+1}^n X_i^{b_{i-k}^X}\\
    \implies \mathtt E_i &= U (I_{k} \otimes e_i^{R,X}) U^\dagger\\
    &= \prod_{i=k+1}^n (U X_i U^\dagger)^{b_{i-k}^X},
\end{align}
and
\begin{align}
    \mathtt S_i^u &:= I_{k} \otimes e_i^{R,Z}\\
    &= \prod_{i=k+1}^n Z_i^{b_{i-k}^Z}\\
    \implies \mathtt S_i &= U (I_{k} \otimes e_i^{R,Z}) U^\dagger\\
    &= \prod_{i=k+1}^n (U Z_i U^\dagger)^{b_{i-k}^Z}.
\end{align}
where $b_{i-k}^X$ and $b_{i-k}^Z$ are the $X$ and $Z$ components of $\vb e_i^{R,X}$ and $\vb e_i^{R,Z}$, respectively.
As this procedure is deterministic, we obtain unique components $\mathtt E_i, \mathtt S_i$, and $\mathtt L_i$ associated with the error pattern $e_i$.

Regarding runtime complexity, the method presented in \cref{sec:function_F_E} requires just $\order{n-k}$ steps per error $E_i$, in order to assemble $\mathtt E_i$ from the precomputed $Z_{h_i}$ and $X_{h_i-n}$ terms. The method presented in  \cref{sec:function_F_L} is more involved. 
Computing $e_i'$ requires $2n$ steps. Determining its stabilizer component $\mathtt S_i$ requires identifying the pivots in $\vb e_i'$ ($\order{n}$ steps) and then multiplying the constituting stabilizers by $\vb e_i'$. As it is constituted by $\order{n-k}$ stabilizers, and accounting for each takes at most $(2n-(n-k-1))=n+k+1$ steps, the whole stabilizer part takes $\order{(n-k)(n+k+1)}$ steps. For the logical component, there are $\order{2k}$ components in $\vb e_i'$, and the whole each operator takes $(2n-(n-k)-2k+1)=n-k+1$ steps, for a total of $\order{(2k)(n+k-1)}$ steps. 
The full procedure requires
\begin{gather}
    \order{n + (n-k)(n+k+1) + (n-k+1)(2k)}\\ \sim \order{n^2, k^2} \sim \order{n^2}
\end{gather}
steps per error $E_i$ to compute $\mathtt L_i$. The computation for $N_E$ errors requires
\begin{equation}
    \order{N_E n^2}
\end{equation}
steps.

For this simpler technique, the greatest computational expense comes from computing $e_i^u$ for each error $E_i$, as the remaining steps can be precomputed and subsequently applied to all errors. 

To facilitate the calculation, we may precompute the $e^u$ patterns associated with all base errors, and use those to compute the pattern $e_i^u$ for each error $E_i$.

From \cite{aaronson_improved_2004}, simulating a stabilizer circuit (i.e. $U$) with $N$ gates takes $\order{n^2 N}$ steps. Since $U$ is built from $\order{n \log^2 n}$ Clifford gates, we have that the full precomputation associated with the base errors scales as
\begin{equation}
    \order{\N n^3 \log^2 n}.
\end{equation}
The cost of computing $e_i^u$ for each $E_i$ then scales as
\begin{equation}
    \order{n\omega}
\end{equation}
where $\omega$ is the order of the error. The full computation for $N_E$ errors requires
\begin{equation}
    \order{N_E n\omega + \N n^3 \log^2 n}
\end{equation}
steps. For cases where $N_E \gg \N$, the simpler approach may lead to a faster implementation, while, for smaller cases, the main approach is faster, as it doesn't require precomputation.

\section{Analysis of numerical results}\label{sec:P_derivation}

To get a good understanding of the performance of the decoding method, we consider an equidistant range for $h$, and we sample $\p$ using the expression
\begin{equation}
    p = 10^{h}.
\end{equation}
For that reason, most of the fits in this section are performed after applying a logarithmic transformation to both the dependent and independent variables. That is, we prefer to work with $\log(\p)$ than $\p$ directly, as it is more numerically stable.

In this section, we verify that $C(p)$ and $P_{\rm failure}$ scale as
\begin{align}
    C(p) &\simeq e^{-\gamma \N \p}\label{eq:C_fit}\\
    &\simeq 1 - \gamma \N \p\\
    P_{\rm failure}(p) &\simeq 1 - e^{-\mu \p^{\eta}},\text{ for }\eta\in\mathbb N\label{eq:P_failure_fit}\\
    &\simeq \mu \p^{\eta},
\end{align}
for $p\ll 1$. $\mu$ and $\gamma$ are positive real parameters, and $\eta$ is the lowest order that the code cannot fully correct. We analyze these expressions separately in the next subsections.

\subsection{Escaped errors}

We start by considering $P_u(p,1)$. For the noise model in \cref{sec:setup}, we expect that the probability of finding escaped errors will be given by
\begin{align}
    P_u(p,1) &\simeq \sum_{i=1}^\N \mqty(\N \\ i) (\gamma\p)^i (1-\gamma\p)^{\N - i}\\
    &= \sum_{i=1}^\N (-1)^{i+1}\mqty(\N \\ i) (\gamma\p)^i\\
    &= 1 - \sum_{i=0}^\N \mqty(\N \\ i) (-\gamma\p)^i,
\end{align}
where $\gamma$ reflects the fraction of errors that can escape. For $\p\ll 1$, the lower order terms dominate, so we may use the approximation
\begin{equation}
    \mqty(\N \\ i) \simeq \frac{\N^i}{i!},\text{ for }\N \gg i,
\end{equation}
and we have
\begin{align}
    P_u(\p,1) &\simeq 1-\sum_{i=0}^\N \frac{(- \N \gamma\p)^i}{i!}\label{eq:P_u_approx}\\
    &\simeq 1-\sum_{i=0}^\infty \frac{(- \N \gamma\p)^i}{i!}\\
    &= 1-e^{-\gamma\N\p}.
\end{align}
Given \cref{eq:P_u_q,eq:C_p}, we end up with \cref{eq:C_fit},
\begin{equation}
    C(p) \simeq e^{-\gamma \N \p},
\end{equation}
where $\gamma$ is an unknown parameter. Since we are considering $\p \ll 1$, this expression may be further simplified into
\begin{equation}
    C(p) \simeq 1 - \gamma \N \p.
\end{equation}

In order to perform a fit, we consider a more general version of this expression, given by
\begin{equation}
    C(p) \simeq \exp(-\gamma \N \p^{\eta_C}),
\end{equation}
and we fit the function
\begin{align}
    \log_{10}(-\log(C(p))) &= \eta_C \log_{10}(\p) + \log_{10}(\gamma \N)\\
    \implies \log_{10}(-b) &= \eta_C \log_{10}(\p) + \beta,\label{eq:C_fit_b}
\end{align}
where $\beta = \log_{10}(\gamma \N)$ and $b$ stems from \cref{eq:P_fail_C}. We use $\log_{10}$ to keep the figures more legible.

\begin{figure}[t]
    \includegraphics[width=\linewidth]{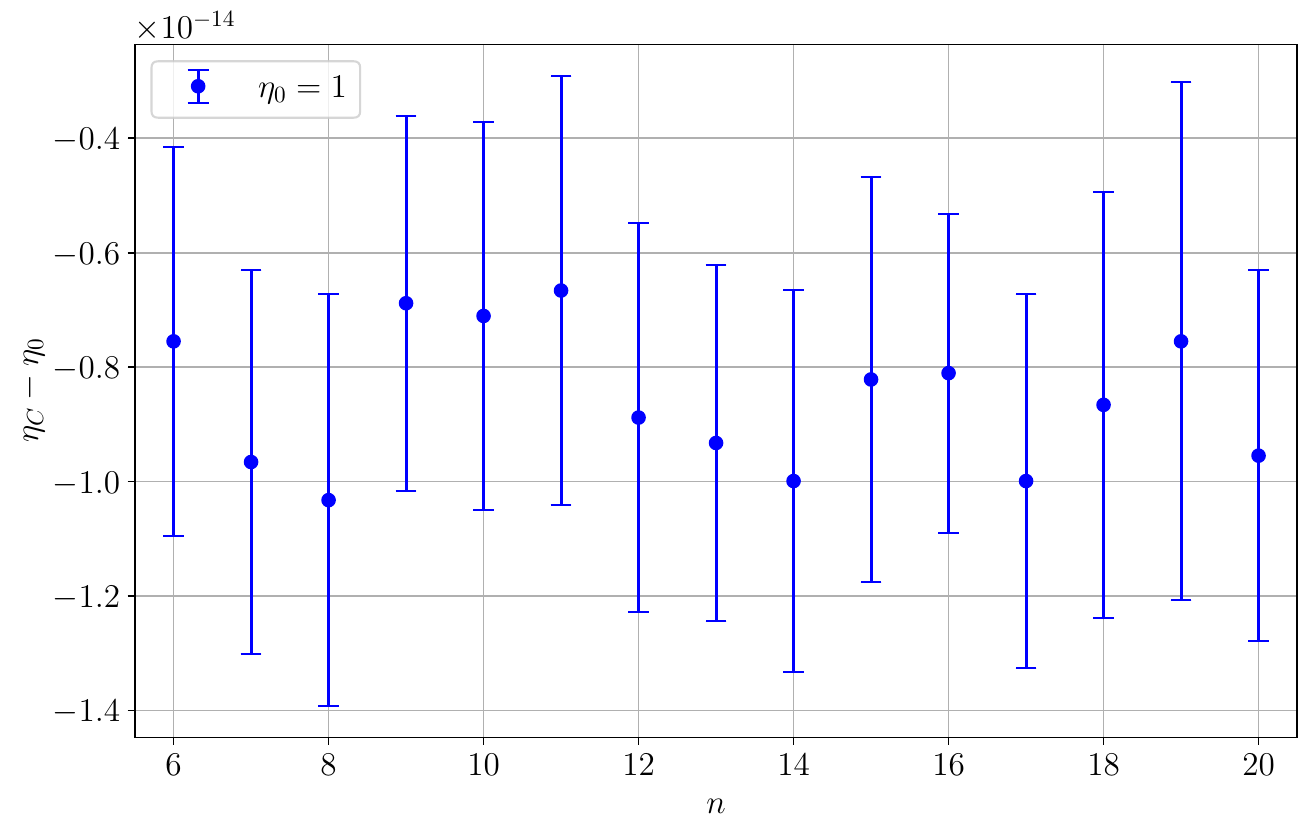}
    \includegraphics[width=\linewidth]{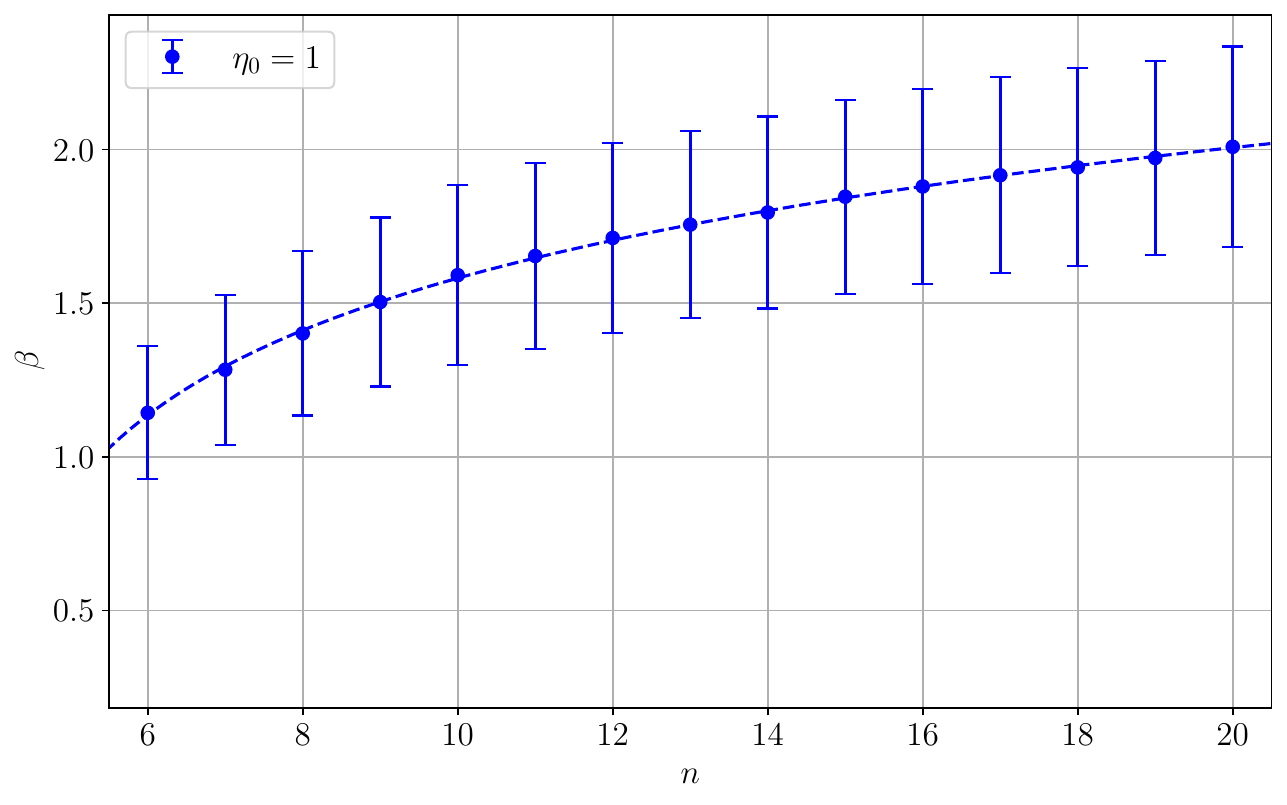}
    \caption{Fit of the various parameters, averaged over 50 random codes for each $n$.}
    \label{fig:fit_parameters_C}
\end{figure}

We may now study the behavior of $\eta_C$ and $\beta$ for different $n$. See \cref{fig:fit_parameters_C} for the results. As expected, we observe that $\eta_C\simeq 1$, regardless of $n$.

From \cref{sec:analytical_threshold}, we have that $\N \sim \order{n^2}$. If we assume that $\gamma \sim \order{\text{poly}(n)}$, then $\beta$ should scale as
\begin{align}
    \beta &= b_1 \log(n - n_\beta) + b_2.
\end{align}
with $\eta_0 = 1$. We assume this expression for its fit. The fitted parameters are given in \cref{tab:parameters}.

\begin{table}[ht]
\centering
\caption{Fitted parameters for $\beta$.}
\label{tab:parameters}
\begin{tabular}{cc}
\toprule
Variable & Value \\
\midrule
$b_1$ & $0.4494$ \\
$n_\beta$ & $3.6453$ \\
$b_2$ & $0.7505$ \\
\bottomrule
\end{tabular}
\end{table}

We may alternatively write $\beta$ as
\begin{equation}
    \beta \simeq \log_{10}(5.630 \cdot (n - 3.6453)^{1.035})
\end{equation}
Given the dependence between $\beta$ and $\N$, we may conclude that
\begin{equation}
    \gamma\N \sim \order{n^{1.035}} \simeq \order{n}, 
\end{equation}
despite the actual $\N$ count scaling with $\order{n^2}$. These results are in line with the expectations from the theoretical analysis of \cref{sec:analytical_threshold}, indicating that we may observe a visible $\pth$.

\subsection{Direct \texorpdfstring{$P_{\rm failure}$}{Pfailure} extrapolation}

We may apply a similar procedure to $P_{\rm failure}$. We empirically observe that \cref{eq:P_failure_fit} holds. However, unlike the previous section, it is no longer the case that $\eta_C \simeq 1$.

If a code is able to correct all errors of order $\omega < \eta$, then we expect $P_{\rm failure}$ to be given by

\begin{align}
    P_{\rm failure} &= \sum_{i=1}^\N f_i(\p, \N) \mqty(\N \\ i) \p^i (1-\p)^{\N - i}\\
    &\simeq \sum_{i=\eta}^\N \tilde f_i \mqty(\N \\ i) \p^i (1-\p)^{\N - i}, \text{ for }\p \ll 1\label{eq:tilde_f_i}\\
    &\simeq \tilde f_\eta \mqty(\N \\ \eta) \p^\eta (1-\p)^{\N - \eta}, \text{ for }\p \ll 1,\\
    &= \mu \p^\eta, \text{ for }\p \ll 1.
\end{align}
for some unknown $\mu$, where $\tilde f_i$ is an approximation of $f_i(\p, \N)$, which is unknown function. In the approximation in \cref{eq:tilde_f_i}, we consider $\tilde f_i$ as a scalar.

\begin{figure}[t]
    \includegraphics[width=\linewidth]{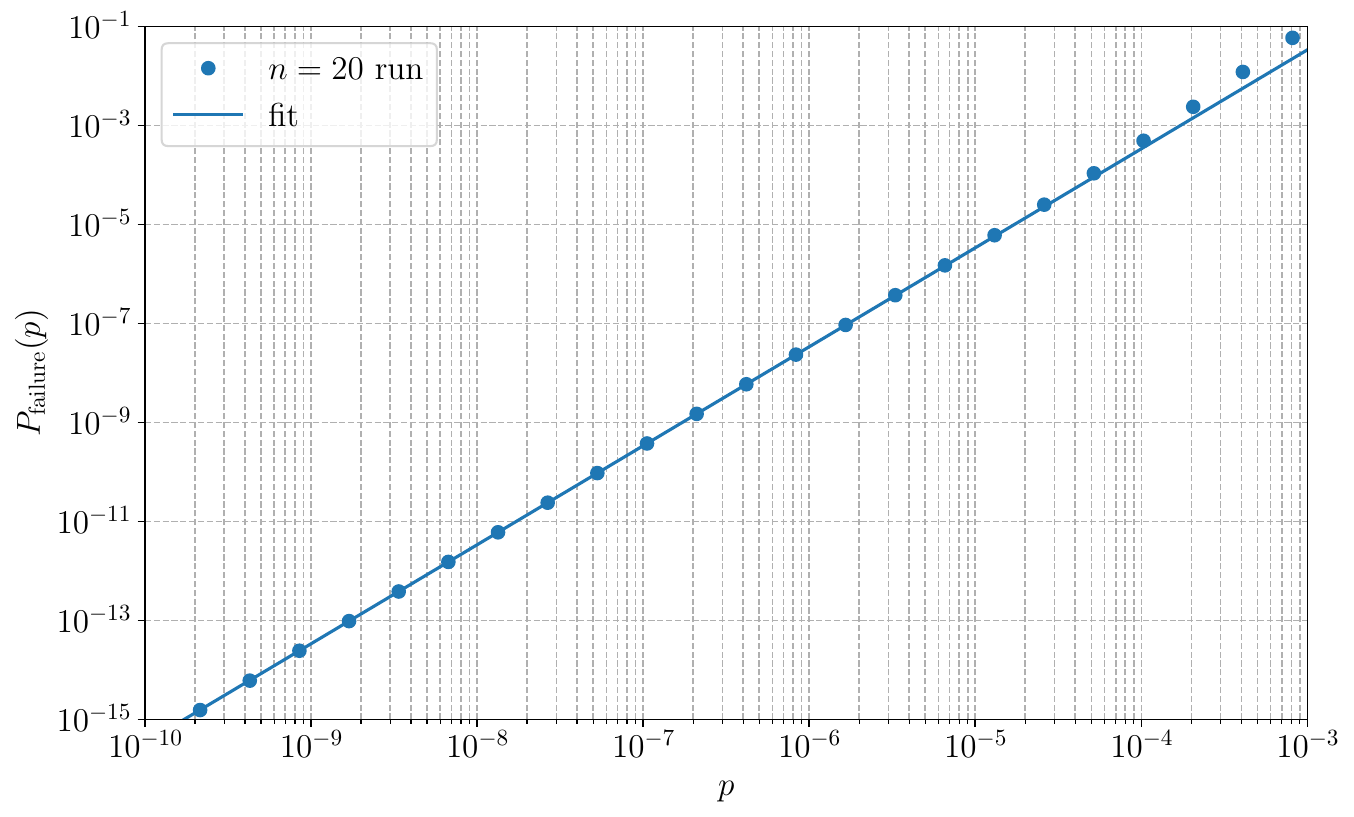}
    \caption{Fit of \cref{eq:C_fit_b}, for a random example with $n=20$. We fit only using values below $10^{-6}$, where it is safe to assume $\p\ll 1$. As expected, the fit worsens for higher $\p$.}
    \label{fig:fit_P}
\end{figure}

In order to perform a fit, we consider an equivalent version of this expression for $\p \ll 1$, given by
\begin{equation}
    P_{\rm failure}(p) \simeq 1 - \exp(-\mu \p^{\eta}),
\end{equation}
and we fit the function
\begin{align}
    \log_{10}(-\log (1-P_{\rm failure})) &= \eta \log_{10}(\p) + \log_{10}(\mu)\\
    &= \eta \log_{10}(\p) + \alpha,\label{eq:P_f_fit}
\end{align}
where $\alpha = \log_{10}(\mu)$. As before, we use $\log_{10}$ to keep the figures more legible. See \cref{fig:fit_P} for an example.

\begin{figure}[t]
    \includegraphics[width=\linewidth]{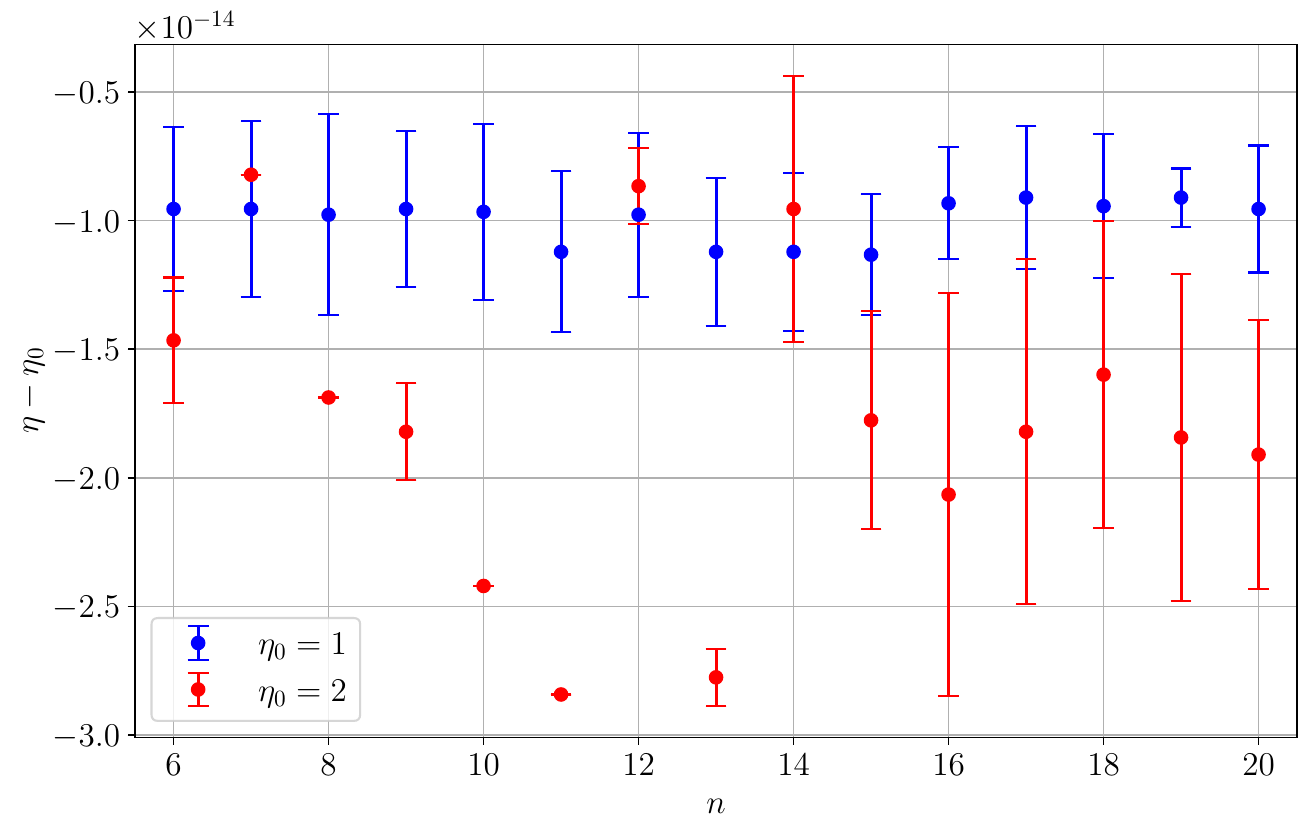}
    \includegraphics[width=\linewidth]{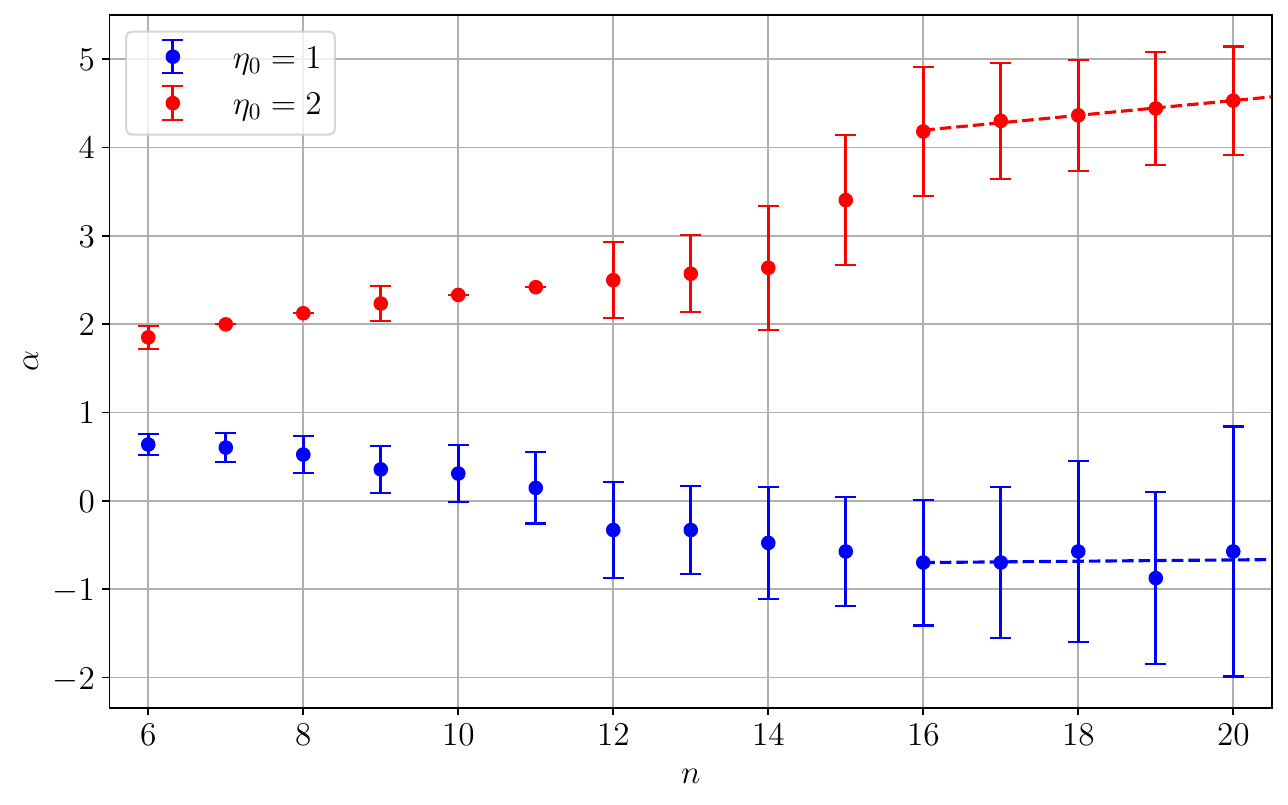}
    \caption{Plot of $\eta$ and $\alpha$. We fit a linear function to $\alpha$ for the higher $n$ values.}
    \label{fig:fit_parameters_P}
\end{figure}

We may now study the behavior of $\eta$ and $\alpha$ for different $n$. See \cref{fig:fit_parameters_P} for the results. In this case, we have different $\eta$ values, and the values cluster around integers. Since the $n$ values we tested are relatively low, we only observe $\eta$ equal to either 1 or 2. As these cases display notably different behavior, we separate their data before fitting.

As before, we fit $\eta$ to an expression of the form
\begin{align}
    \eta &= \eta_0 + d_1 \log(1 + \exp(d_2 \cdot (n - n_P))).
\end{align}
We empirically observe that the $\alpha$ variable is better fitted by a simple linear expression
\begin{equation}
    \alpha = a_1 n + a_2.
\end{equation}

The fitted parameters are given in \cref{tab:alpha}.

\begin{table}[ht]
\centering
\caption{Fitted parameters for $\alpha$.}
\label{tab:alpha}
\begin{tabular}{ccc}
\toprule
Variable & $\eta_0 = 1$ & $\eta_0 = 2$ \\
\midrule
$a_1$ & $-0.0074$ & $0.0838$ \\
$a_2$ & $-0.8170$ & $2.8530$ 
\end{tabular}
\end{table}

\begin{figure}[t]
    \includegraphics[width=\linewidth]{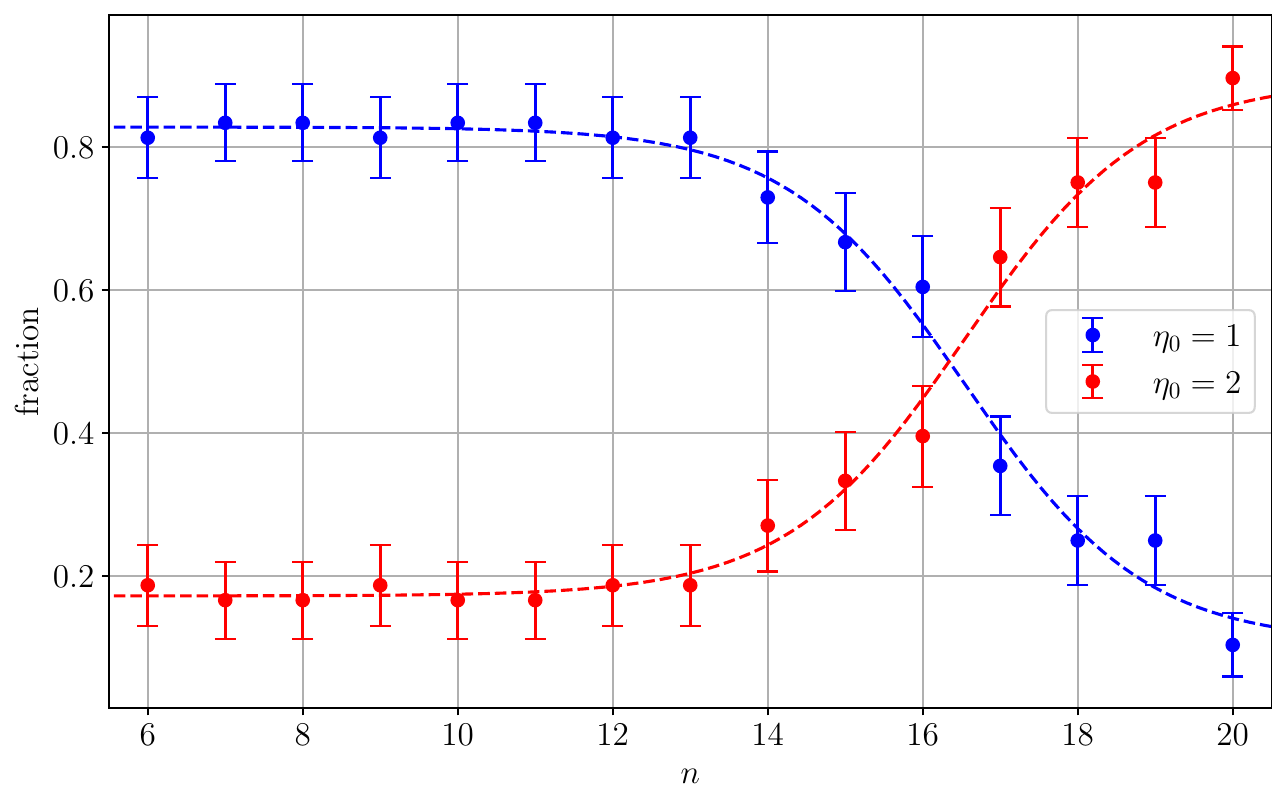}
    \caption{Fraction of the simulations with different $\eta_0$ values.}
    \label{fig:fraction_parameters}
\end{figure}

We may also have a look at the probability of having different $\eta$ values. The empirical results can be seen in \cref{fig:fraction_parameters,tab:fraction_parameters}. We observe that, for low $n$ values, the fraction of random codes with $\eta=1$ and $\eta=2$ is roughly constant, since the code's capabilities are not large enough to generally correct all $\omega=1$ errors. As $n$ increases, the probability that the code corrects all order-1 errors also increases, and we so the fraction of $\eta=2$ cases increases. Its behavior follows a sigmoid-like function, of the form
\begin{equation}
    s_1\qty(\frac{1}{1 + e^{-s_2(n-n_f)}} - \frac{1}{2}) + s_3.
\end{equation}
The errorbars indicate the standard deviation, which stems from the finite number (48) of samples taken for each $n$ value. Also, note the symmetry in the parameters in \cref{tab:fraction_parameters}, reflecting the fact that the fractions must add up to one.

\begin{table}[htbp]
\centering
\caption{Fitted parameters.}
\label{tab:fraction_parameters}
\begin{tabular}{ccc}
\toprule
Variable & $\eta_0 = 1$ & $\eta_0 = 2$ \\
\midrule
$s_1$ & $-0.7206$ & $0.7206$ \\
$s_2$ & $0.8665$ & $0.8665$ \\
$n_f$ & $16.5506$ & $16.5506$ \\
$s_3$ & $0.4673$ & $0.5327$ 
\end{tabular}
\end{table}

For very low $n \leq 12$, we observe that about $20\%$ of the codes have $\eta=2$. On further inspection, their seemingly high performance does not stem from strong correction capabilities, but from the fact that, due to random chance, the stabilizers of these codes have relatively low weight, leading to a very low number of CNOT gates in the syndrome extraction circuit, and consequently fewer errors needed to consider. In a more complex setting where there are additional sources of error, we would expect these codes in particular to perform poorly.

\subsection{Extrapolating behavior for larger \texorpdfstring{$n$}{n}}

We may take the results above and use them to extrapolate the performance for larger $n$ values. We also incorporate the uncertainty observed in the numerical data by using the standard deviation observed for the sampled $n$ to estimate the deviation for larger $n$.

\section*{Acknowledgments}
All authors are thankful to Dr. Bill Munro (NTT Basic Research Labs, Japan) and Prof. Kae Nemoto (National Institute of Informatics, Japan) for insightful discussions about QGRAND.
Francisco Monteiro is grateful to Prof. Frank Kschischang (University of Toronto) for discussions on noise-guessing decoding, and to Dr. Ioannis Chatzigeorgiou (Lancaster University) for discussions on random linear codes and noise-guessing decoding. 

\bibliographystyle{IEEEtran}
\bibliography{FT_paper_filtered.bib}

\vfill\eject
%\vskip -5pt plus -5 fill
\begin{IEEEbiography}[{\includegraphics[width=1in,height=1.25in,clip,keepaspectratio]{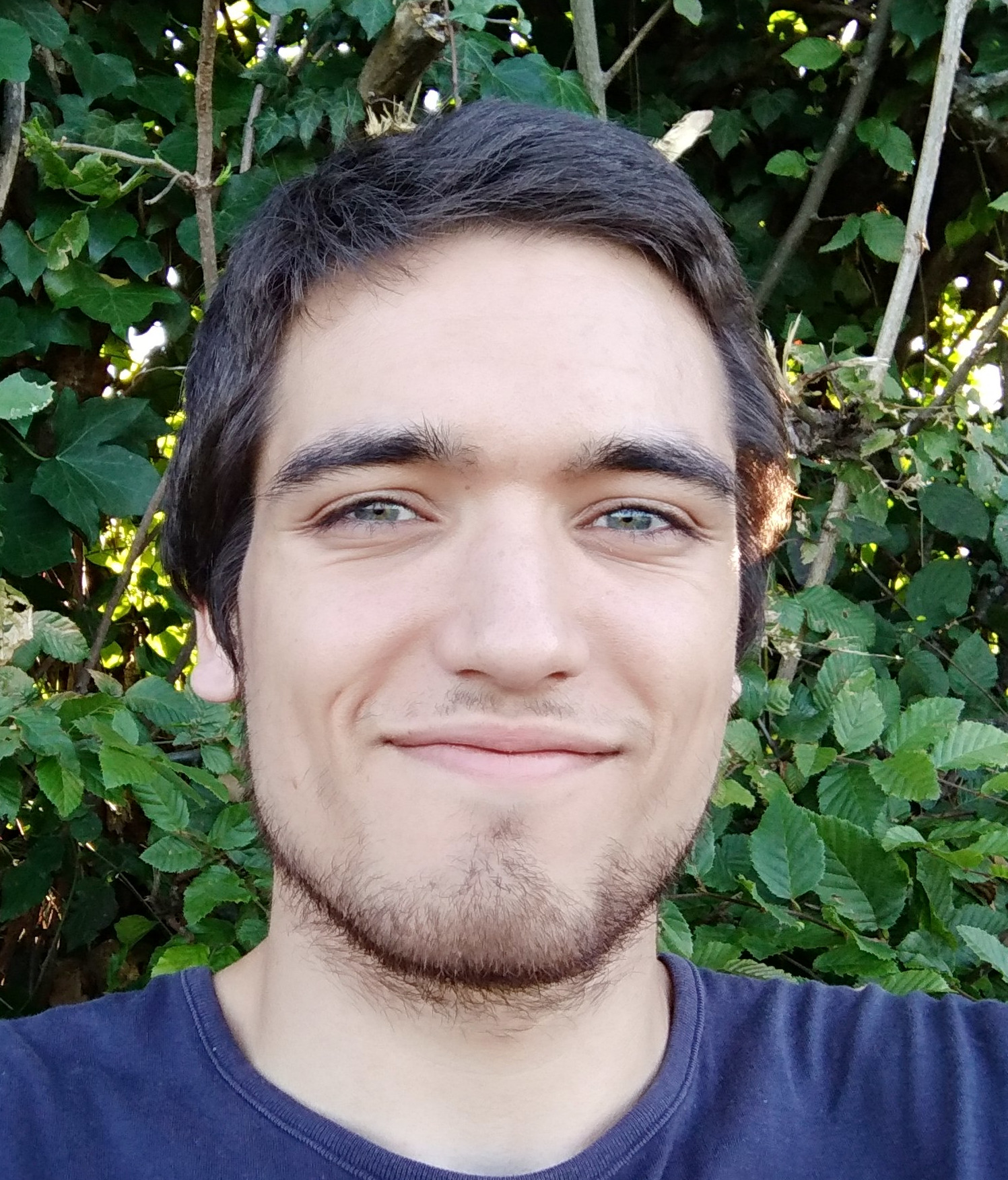}}]{Diogo Cruz} obtained his BSc and MSc degrees in Physics Engineering from Instituto Superior Técnico (IST), University of Lisbon, Portugal. He is currently a PhD student in Physics, also at IST, and is a researcher at Instituto de Telecomunicações, Lisbon, Portugal. He was a Calouste Gulbenkian Scholar in 2018/2019.
\end{IEEEbiography}

%\vskip 5pt plus -5 fill
\vskip -50pt
\begin{IEEEbiography}[{\includegraphics[width=1in,height=1.25in,clip,keepaspectratio]{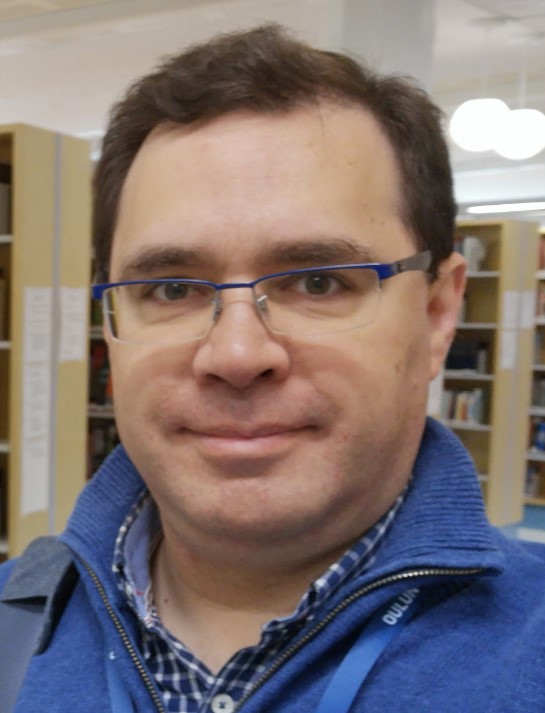}}]{Francisco A. Monteiro} (M'07) is Associate Professor in the Dep. of Information Science and Technology at Iscte - University Institute of Lisbon, and a researcher at Instituto de Telecomunicações, Lisbon, Portugal. He holds a PhD from the University of Cambridge, UK, and the Licenciatura and MSc degrees in Electrical and Computer Engineering from IST, University of Lisbon, where he also became a Teaching Assistant. He held visiting research positions at the Universities of Toronto (Canada), Lancaster (UK), Oulu (Finland), and Pompeu Fabra (Barcelona, Spain). He has won two best paper prizes awards at IEEE conferences (2004 and 2007), a Young Engineer Prize (3rd place) from the Portuguese Engineers Institution (Ordem dos Engenheiros) in 2002, and for two years in a row was a recipient of Exemplary Reviewer Awards from the IEEE Wireless Communications Letters (in 2014 and in 2015). He co-edited the book ``MIMO Processing for 4G and Beyond: Fundamentals and Evolution'', published by CRC Press in 2014. In 2016 he was the Lead Guest Editor of a special issue on Network Coding of the EURASIP Journal on Advances in Signal Processing. He was a general chair of ISWCS 2018 - The 15th International Symposium on Wireless Communication Systems, an IEEE major conference in wireless communications.
\end{IEEEbiography}

\vskip -50pt
\begin{IEEEbiography}[{\includegraphics[width=1in,height=1.25in,clip,keepaspectratio]{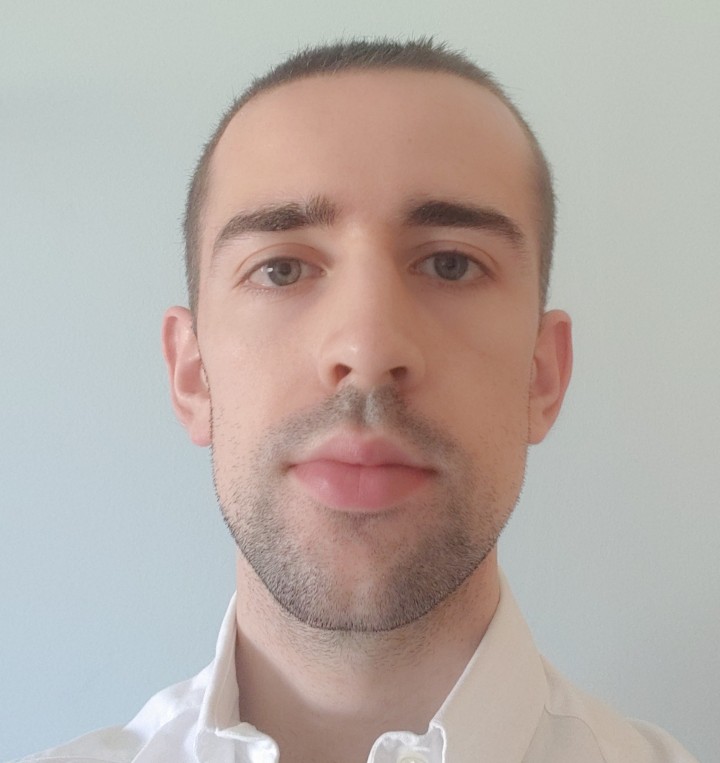}}]{André Roque} obtained both his BSc. and MSc. degrees in Applied Mathematics and Computation, from Instituto Superior Técnico (IST, University of Lisbon), Portugal in 2023. Since 2023 he is with the Physics of Information and Quantum Technologies Group, at Instituto de Telecomunicações as a Research Assistant.
\end{IEEEbiography}

%\vskip 0pt plus -1fil
\vskip -50pt
\begin{IEEEbiography}[{\includegraphics[width=1in,height=1.25in,clip,keepaspectratio]{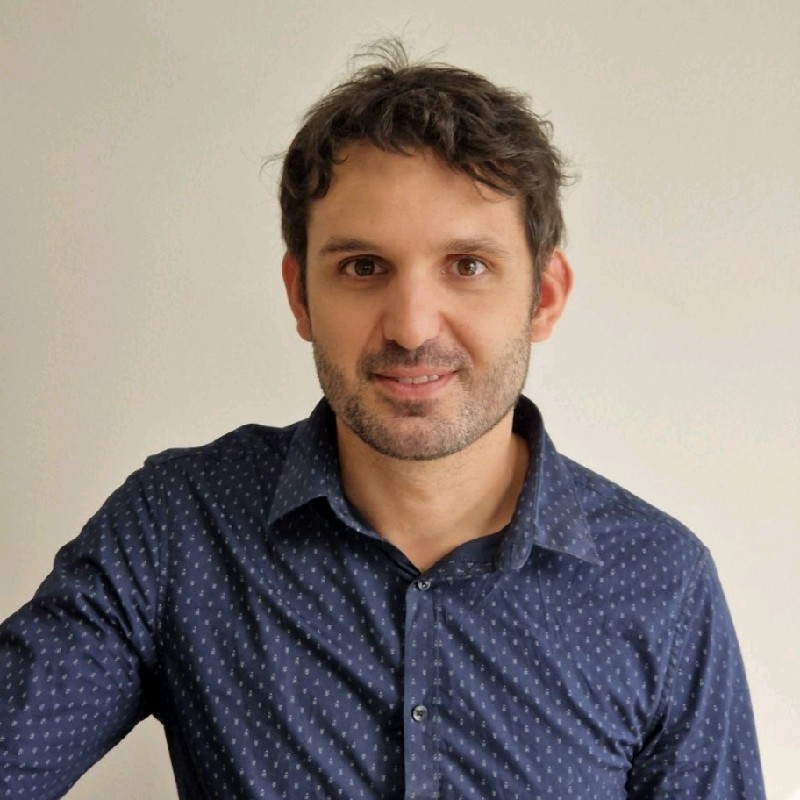}}]{Bruno C. Coutinho} obtained his PhD in Physics from Northeastern University, USA, in 2016, and both his BSc. and MSc. in Physics, from the University of Aveiro, Portugal, in 2009 and 2011, respectively. Since 2017 he is with the Physics of Information and Quantum Technologies Group, at Instituto de Telecomunicações, initially as a postdoc and later as a Research Fellow.
\end{IEEEbiography}

\end{document}